\documentclass{aastex} 
\usepackage{emulateapj5, apjfonts, psfig, epsfig} 
\begin{document}

\slugcomment{AJ, in press} 
 
\title{The Dwarf Spheroidal Companions to M31:  Variable Stars in Andromeda~VI\footnotemark} 
\footnotetext{Based on observations with the NASA/ESA {\it Hubble Space Telescope}, 
obtained at the Space Telescope Science Institute, which is operated by the 
Association of Universities for Research in Astronomy, Inc.\, (AURA), under 
NASA Contract NAS 5-26555.}

\shorttitle{And~VI Variable Stars}
\shortauthors{Pritzl et al.}

\received{}
\revised{}
\accepted{}
 
\author{Barton J. Pritzl and Taft E. Armandroff} 
\affil{National Optical Astronomy Observatory, P.O.\ Box 26732, Tucson, AZ 
      85726} 
\email{pritzl@noao.edu, armand@noao.edu} 

\author{George H. Jacoby} 
\affil{WIYN Observatory, P.O.\ Box 26732, Tucson, AZ, 85726} 
\email{gjacoby@noao.edu}

\and
\author{Gary S. Da Costa} 
\affil{Research School of Astronomy and Astrophysics, Institute of 
       Advanced Studies, The Australian National University, Cotter Road, 
       Weston, ACT 2611, Australia} 
\email{gdc@mso.anu.edu.au}

\begin{abstract}

We have surveyed Andromeda~VI, a dwarf spheroidal galaxy companion to M31, 
for variable stars using F450W and F555W observations obtained with the 
{\it Hubble Space Telescope}.  A total of 118 variables were found, with 111 
being RR~Lyrae, 6 anomalous Cepheids, and 1 variable we were unable to 
classify.  We find that the Andromeda~VI anomalous Cepheids have properties 
consistent with those of anomalous Cepheids in other dwarf spheroidal 
galaxies.  We revise the existing period--luminosity relations for these 
variables.  Further, using these and other available data, we show that there 
is no clear difference between fundamental and first-overtone anomalous 
Cepheids in a period-amplitude diagram at shorter periods, unlike the RR~Lyrae.  
For the Andromeda~VI RR Lyrae, we find that they lie close to the Oosterhoff 
type~I Galactic globular clusters in the period-amplitude diagram, although 
the mean period of the RRab stars, $\langle P_{ab} \rangle = 0.588$d, is 
slightly longer than the typical Oosterhoff type~I cluster.  The mean $V$ 
magnitude of the RR~Lyrae in Andromeda~VI is $25.29\pm0.03$, resulting in a 
distance $815\pm25$~kpc on the Lee, Demarque, \& Zinn distance scale.  This is 
consistent with the distance derived from the $I$ magnitude of the tip of the 
red giant branch.  Similarly, the properties of the RR~Lyrae indicate a mean 
abundance for Andromeda~VI which is consistent with that derived from the mean 
red giant branch color.

\end{abstract}


\section{Introduction} 

Galactic dwarf spheroidal galaxies (dSphs) have been shown to be quite 
diverse in their star formation histories (e.g., Da~Costa 1998; Grebel 
1999).  These galaxies generally also possess a range in metallicity among 
their stars.  A detailed examination of these and other dwarf galaxies is 
important in understanding galaxy formation and gives insight into 
cosmology (e.g., Klypin et al.\ 1999).  The diverse stellar populations in 
these systems is also reflected in their variable stars.  dSphs contain 
significant numbers of RR Lyrae (RRL) implying an older ($t > 10$~Gyr) 
stellar population.  Contrary to Galactic globular clusters whose RRab stars 
can be classified into one of two Oosterhoff types (Oosterhoff 1939), the 
mean period of the RRab stars in a number of dSphs is ``intermediate" between 
the two types.  A second difference between Galactic globular clusters and 
dSphs as regards their variable star content, is the existence of anomalous 
Cepheids (ACs) in dSphs.  Only one AC is known among the entire Galactic 
globular cluster population, V19 in NGC 5466 (Zinn \& King 1982), but 
there exists at least one AC in every dSph surveyed for variable stars.  
These fundamental differences between dSphs and Galactic globular clusters are 
presumably indicative of the differences in the stellar populations of these 
systems.  

Clearly, a large and diverse sample of variable stars in dSphs can contribute 
to the understanding of the underlying stellar populations in these systems.  
While there have been many detailed variability surveys for the Galactic 
dSphs (e.g., Kaluzny et al.\ 1995; Mateo, Fischer, \& Krzeminski 1995; 
Siegel \& Majewski 2000; Held et al.\ 2001; Bersier \& Wood 2002), there are 
none available for the M31 dSph 
companions.  In this paper we examine the variable star content of the M31 
dSph Andromeda~VI (And~VI, also known as the Pegasus dSph).  Armandroff, 
Jacoby, \& Davies (1999, hereafter AJD99) found And~VI to have a mean 
metallicity of $\langle {\rm [Fe/H]} \rangle = -1.58\pm0.20$~dex from the mean 
($V$--$I$) color of the red giant branch.  This places And~VI among the more 
metal--rich of the dSphs within the Local Group.  AJD99 also derived a 
distance of $775\pm35$~kpc for And~VI from the $I$ magnitude of the tip of 
the red giant branch.  A detailed analysis of the And~VI color--magnitude diagram, 
based on HST/WFPC2 images, will be presented in Armandroff et al.\ (2002).  
Here, we present the discovery of 118 variables in And~VI from that data; 
111 RRL, 6 ACs, and 1 whose classification is uncertain.  Light curves and mean 
properties are given for each variable.  We then use these data to compare 
the properties of the And~VI variables with those for other dSphs.

\section{Observations and Reductions} 

As part of our GO program 8272, the {\it Hubble Space Telescope} imaged And~VI 
with the WFPC2 instrument on 1999 October 25 and, with the same orientation, on 
1999 October 27.  For each set of observations, four 1100s exposures through 
the F555W filter and eight 1300s exposures through the F450W filter were taken.  
The second set of observations was offset slightly from the first set in order 
to aid in distinguishing real stars from image defects.  The 
images were taken with a gain of 7 electrons/ADU.\@  The raw frames were 
processed by the standard STScI pipeline.  
Each frame was separated into individual images for each CCD and the 
vignetted areas were trimmed using the limits defined in the WFPC2 Handbook.

\subsection{Photometry} 

In order to create a star list for photometry that is relatively free of 
contamination by cosmic ray events, star detection was performed on a cleaned 
image.  To do this, we used the Peter Stetson routine, {\sc montage ii}.  
This program creates a median image from all available images.  The advantage 
of this is that the median image eliminates nearly all of the cosmic rays and 
other defects that are found on the individual images.  A search for stellar 
objects was then performed on the median image for each CCD using a full-width 
at half maximum of 1.6 pixels.  Aperture photometry on these median images 
was then obtained using Stetson's (1992) stand-alone version of 
{\sc daophot ii}.\@  An aperture radius of 2.0 pixels was used.  The 
{\sc allstar} routine was then employed to obtain profile--fitting photometry 
for the stars on the median images, adopting a fitting radius of 1.6 pixels.  
The point-spread functions for each CCD were obtained from Peter Stetson; they 
were created for reducing data in the {\it Extragalactic Distance Scale Key 
Project} (Stetson et al.\ 1998).  The objects that were fit by {\sc allstar} 
on the median image were assumed to be stars and not image defects or resolved 
galaxies. 

The master list of objects was then used by {\sc allframe} (Stetson 1994) 
to obtain profile-fitting photometry for each individual CCD exposure.  This 
program is designed to reduce all of the images simultaneously for each CCD.\@  
The resulting photometry was investigated for variable stars using the 
Stetson routine {\sc daomaster}.  {\sc daomaster} compared the rms scatter 
in the photometric values to that expected from the photometric errors 
returned by the {\sc allframe} program.  The PC was searched for variable 
stars and only turned up a handful of candidate RRL with no ACs.  
Since this small number of RRL would add little to the final results, 
we decided to ignore the PC data.

\subsection{CTE and Aperture Corrections} 

It is well known that the WFPC2 CCDs suffer from poor charge-transfer 
efficiency (CTE) which affects the photometry (Holtzman et al.\ 1995; Stetson 
1998; Whitmore, Heyer, \& Casertano 1999; Dolphin 2000).  In order to correct 
for this effect we applied equations (2c) and (3c) from Whitmore, Heyer, \& 
Casertano (1999) to the profile-fitting photometry for each frame.

To place the photometry on the system of Holtzman et al.\ (1995), aperture 
corrections are required to convert the profile--fit magnitudes to 
magnitudes within a 5 pixel radius aperture.  As part of this process, a 
comparison of the profile--fit photometry for each individual frame to 
that for an adopted ``reference" frame (the first image in the sequence 
for the first pointing) was carried out.  This revealed that 
there were cyclic differences from frame--to--frame that followed the exposure 
sequence of the observations.  As a result, we decided to determine the 
aperture correction for the adopted ``reference" frame and base the aperture 
corrections for the other frames on the differences seen between the 
profile-fitting photometry of each frame as compared to the ``reference" frame.  

In order to do this, the difference between the 5 pixel radius aperture 
magnitudes from the ``gcombined" frames\footnote{The gcombined frames were 
used because they have a higher signal-to-noise ratio and cosmic rays are 
eliminated.} for each CCD (Armandroff et al.\ 2002) 
and the profile--fit magnitudes for the ``reference" frame was then computed 
for those stars that had no neighbors within a 7 pixel radius.  We then 
generated the weighted mean of these differences for all stars brighter than 
the horizontal branch by approximately 1 magnitude for each filter.  In this 
calculation, stars whose differences were greater than 0.4~mag were excluded.  
This gave us the aperture correction for the ``reference" frame.  The aperture 
corrections for the rest of the frames were then calculated from 
the aperture correction for the ``reference" frame and the differences in 
the profile--fit photometry between this frame and each individual frame.    

The resulting aperture corrections were then applied 
to the profile--fit photometry along with the CTE corrections.  After 
correction, the resulting mean 
differences in the photometry between the frames were, for the most part, 
below 0.01~mag with the maximum difference being around 0.015~mag.

\begin{figure*}[t]
  \centerline{\psfig{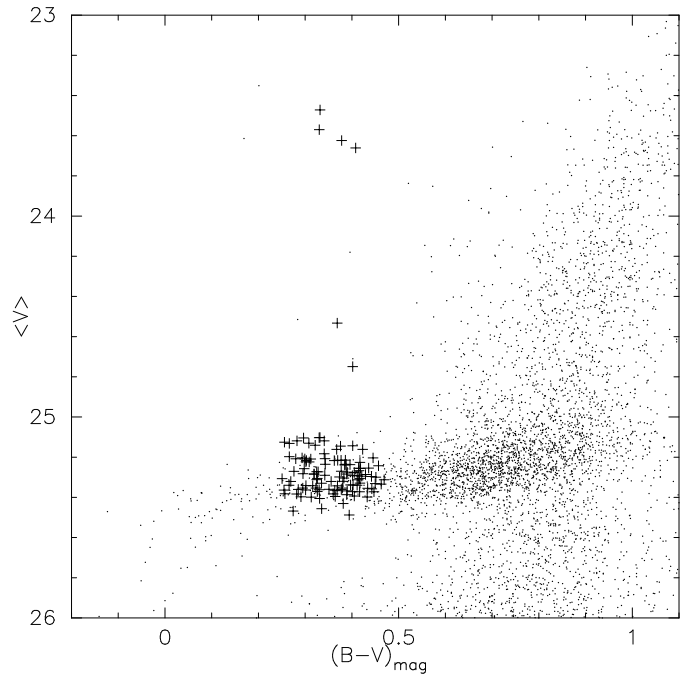}}
  \caption{And~VI color-magnitude diagram showing the location of the 
           RR~Lyrae and anomalous Cepheids marked as crosses.} 
  \label{Fig01} 
\end{figure*}

\section{Light Curves} 

With at most 16 epochs in F450W and 8 epochs in F555W, phasing the photometry 
and finding accurate periods in both colors presents a challenge.  In order 
to employ the maximum number of photometric measures in the period finding, 
we sought to use the F450W and F555W data together.  Therefore, we need to 
determine the magnitude offset for each star to place the F555W observations 
on the F450W magnitude scale.  We began this process by noting that the last 
F555W observation and the first F450W observation were taken consecutively.  
Therefore, we can assume that this F555W magnitude, which we denote by 
$m_{0{\rm, F555W}}$, corresponds to the F450W magnitude, which we denote by 
$m_{0, {\rm F450W}}$.  The other F555W magnitudes are then converted to their 
F450W equivalents by using the equation: 

\begin{equation}
m_{{\rm n, F450W}} = m_{{\rm 0, F450W}} + 1.3(m_{{\rm F555W}} - m_{{\rm 0, F555W}}) 
\end{equation} 
 
\noindent 
where $m_{{\rm n, F450W}}$ is the F450W magnitude derived from the F555W 
magnitude $m_{{\rm F555W}}$.  The 1.3 
value is the approximate ratio of the amplitudes in $B$ to $V$ for the light 
curves.  This value was determined from a number of RRL in various Galactic 
GCs which had clean light curves in both colors.  While this ratio may not be 
exactly the same as the F450W to F555W ratio for the amplitudes, a difference 
of $\pm 0.1$ in the amplitude ratio does not affect 
the results.  This combined dataset was then used to determine the period of 
the star through the use of routines created by Andrew Layden\footnote{Available 
at http://physics.bgsu.edu/$\sim$layden/ASTRO/DATA/EXPORT/lc\_progs.htm from 
A. C. Layden.} (Layden \& Sarajedini 2000 and references therein).  The 
program is designed to determine the most likely period from the chi--squared 
minima by fitting the photometry of the variable star with 10 templates over 
a selected range of periods.  Since there is little difference in light 
curve shape between RRL and ACs, we are confident that the templates will work 
well for any ACs we find.  Due to the aliasing from the observations, 
these periods were tested in a period--amplitude diagram to determine their 
accuracy.  A small number of RRL with periods scattered from where the majority 
were found in this diagram were revised in order to reduce the scatter.  
These typically were stars that had 
large gaps in their light curves making it difficult to accurately determine 
their period.  The combination of the template-fitting program and the 
period-amplitude diagram allowed us to reduce the likelihood that the given 
period for a variable is an alias of the true period, although there is a 
slight possibility that a small number of the And~VI variables may have an 
alias period.  The typical uncertainty in the periods is about $\pm0.005$~day. 

Once an accurate period was determined, we fit a light curve template to the 
combined data by another routine created by Andrew Layden.  A copy of this 
template was then converted back to the F555W system in a reverse of 
Eq.\ 1.  Having a template for each filter now allows us to make 
use of Eq.\ 8 and the coefficients in Table~7 of Holtzman et al.\ (1995) 
to calibrate the F450W,F555W templates to $B$,$V$ templates for each phase 
along the light curve.  The individual F450W,F555W data points at each 
phase were converted to $B$,$V$ through the color information provided by 
the template $B$,$V$ light curves.  New template light curves were then 
fit to the $B$,$V$ data points.  The preceding two steps were repeated until 
convergence of the $B$,$V$ magnitudes was achieved.

In the following sections, we use the intensity-weighted mean $\langle V 
\rangle$, $\langle B \rangle$ magnitudes and magnitude-weighted ($\bv$) colors 
for each variable, determined by a spline fit to the $B$,$V$ light curve 
templates.

\begin{figure*}[t] 
  \centerline{\psfig{figure=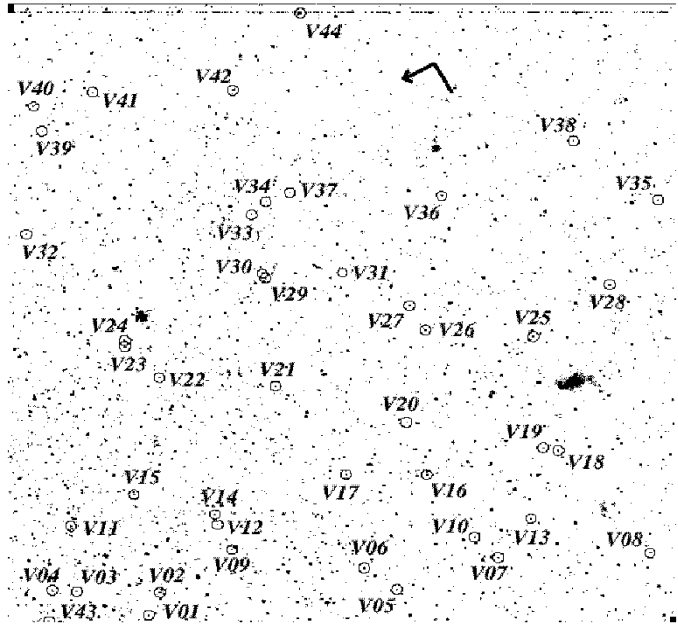,height=3.25in,width=3.50in}} 
  \caption{Finding charts for the And~VI variable stars.  The WFC2 
           ($1.2\arcmin$x$1.3\arcmin$), WFC3 ($1.3\arcmin$x$1.2\arcmin$), and 
           WFC4 ($1.2\arcmin$x$1.2\arcmin$) images are each shown in a panel.  
           North and east directions are shown with the arrow pointing toward 
           the north.} 
  \label{Fig02}
\end{figure*} 

\begin{figure*}[t] 
  \figurenum{2 cont}
  \centerline{\psfig{figure=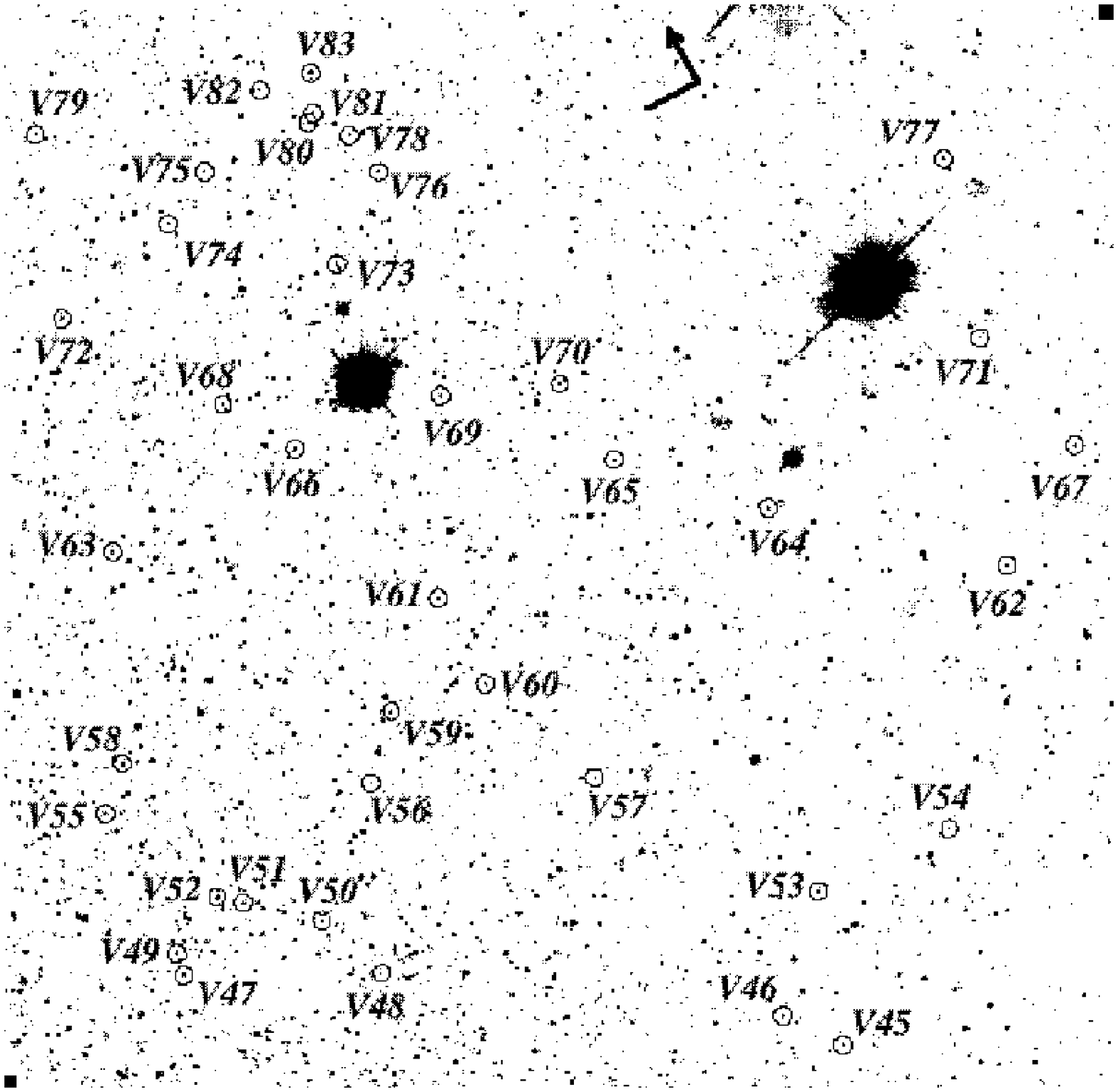,height=3.25in,width=3.50in}} 
  \caption{Finding charts for the And~VI variable stars.  The WFC2 
           ($1.2\arcmin$x$1.3\arcmin$), WFC3 ($1.3\arcmin$x$1.2\arcmin$), and 
           WFC4 ($1.2\arcmin$x$1.2\arcmin$) images are each shown in a panel.  
           North and east directions are shown with the arrow pointing toward 
           the north.} 
  \label{Fig02}
\end{figure*} 

\begin{figure*}[t] 
  \figurenum{2 cont}
  \centerline{\psfig{figure=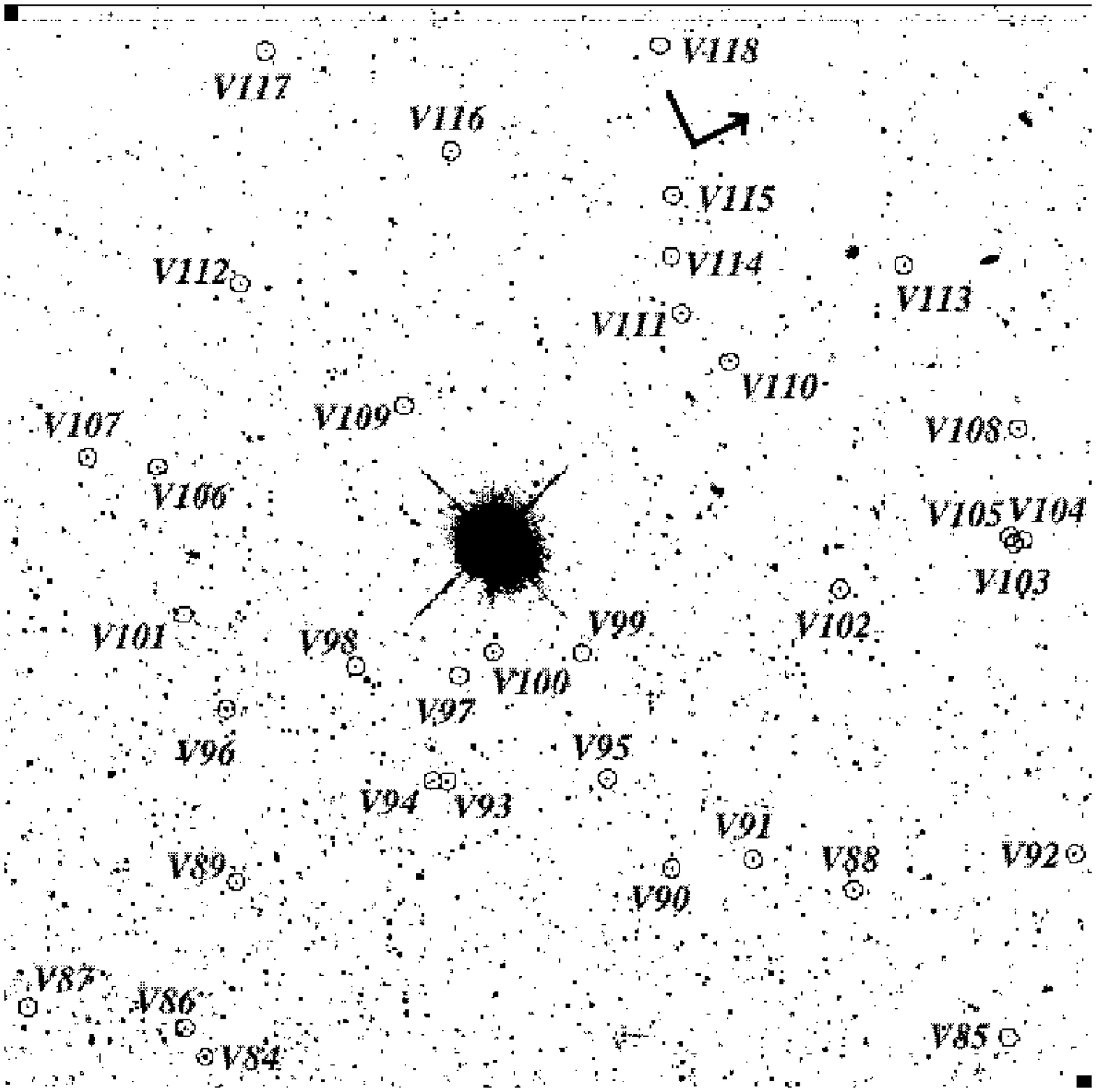,height=3.25in,width=3.50in}} 
  \caption{Finding charts for the And~VI variable stars.  The WFC2 
           ($1.2\arcmin$x$1.3\arcmin$), WFC3 ($1.3\arcmin$x$1.2\arcmin$), and 
           WFC4 ($1.2\arcmin$x$1.2\arcmin$) images are each shown in a panel.  
           North and east directions are shown with the arrow pointing toward 
           the north.} 
  \label{Fig02}
\end{figure*}

\subsection{Variable Star Colors} 

Choosing the correct magnitude offset as discussed in \S3 provided a challenge 
with the variable stars due to the varying brightness along the light curve.  
However, for RRL with uniform reddening, it is known that the colors during 
minimum light are approximately the same.  With this in mind, we calibrated 
those stars for which we had photometry in both filters during minimum 
light.  These variable stars fell within the expected instability strip, 
which is approximately $0.26<(\bv)<0.46$ given $E(\bv)=0.06$ from AJD99.  
Knowing what the magnitude offsets should look like from these variable 
stars, we were able to accurately calibrate the approximately 25\% of the 
variable stars which only had photometry during minimum light for one filter. 

In Figure~1 we show the location of the variables within the color-magnitude 
diagram.  All variables lie within the instability 
strip with the RRL forming a distinct group along the horizontal branch, while 
the brighter stars are likely to be ACs.  We investigated the stars scattered 
about the location of the ACs and found them to show no variability.  
It should be noted that the colors and magnitudes for all the stars have 
inherent uncertainties of order 0.02-0.03 mag due to the method we have used to 
calibrate the data to the $B$,$V$ system.  Nevertheless, our approach is 
adequate to investigate the general properties of And~VI as 
demonstrated in the following sections.

In Table~1 we present the photometric properties for the 118 variables while 
their photometric $B$ and $V$ data are in Tables~2 and 3.  Column~1 of Table~1 
lists the star's ID, while the next two columns give the RA and Dec.  Finding charts 
for the variables are found in Figure~2.  Of these variables, 6 are likely 
ACs and they are discussed in the following section with their light curves 
shown in Figure~3.  A further 111 are 
RRL and Figure~4 shows their light curves with their properties discussed 
in \S5, with the exception of V103.  The variable V103 only had photometry 
along the descending branch of the light curve.  Therefore no templates 
adequately fit the data.  We were unable to 
classify one candidate variable (V34).  The best fit for the data 
implies that it is a contact binary.  On the other hand, its magnitude and 
color place it among the RRL.  For this reason we have not definitively 
classified this variable and have left it out of our analysis.

\begin{figure*}[t] 
  \centerline{\psfig{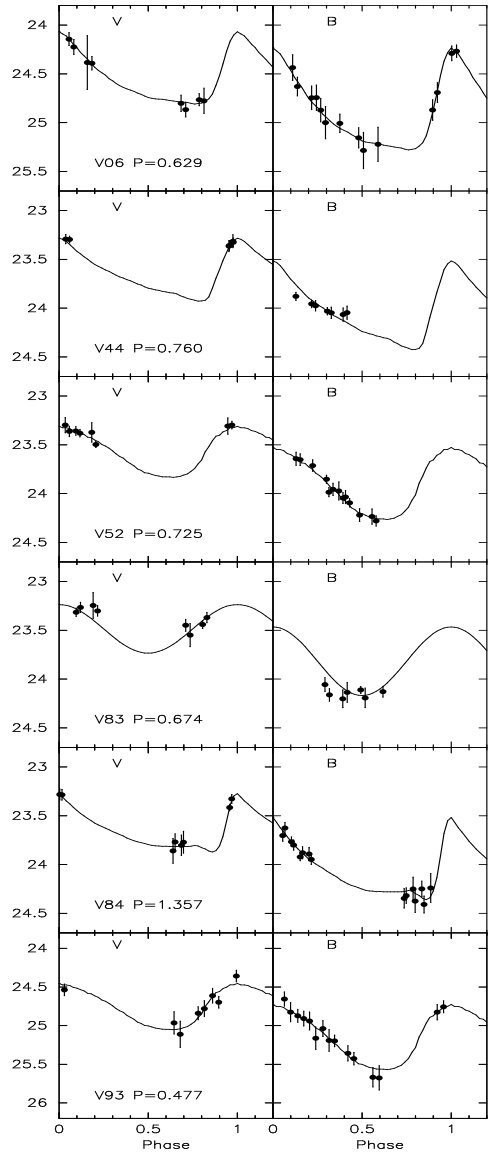}} 
  \caption{And~VI anomalous Cepheid light curves with the observations shown 
            as filled circles and the fitted template displayed as a curve.} 
  \label{Fig03} 
\end{figure*}   

\begin{figure*}[t]
  \centerline{\psfig{figure=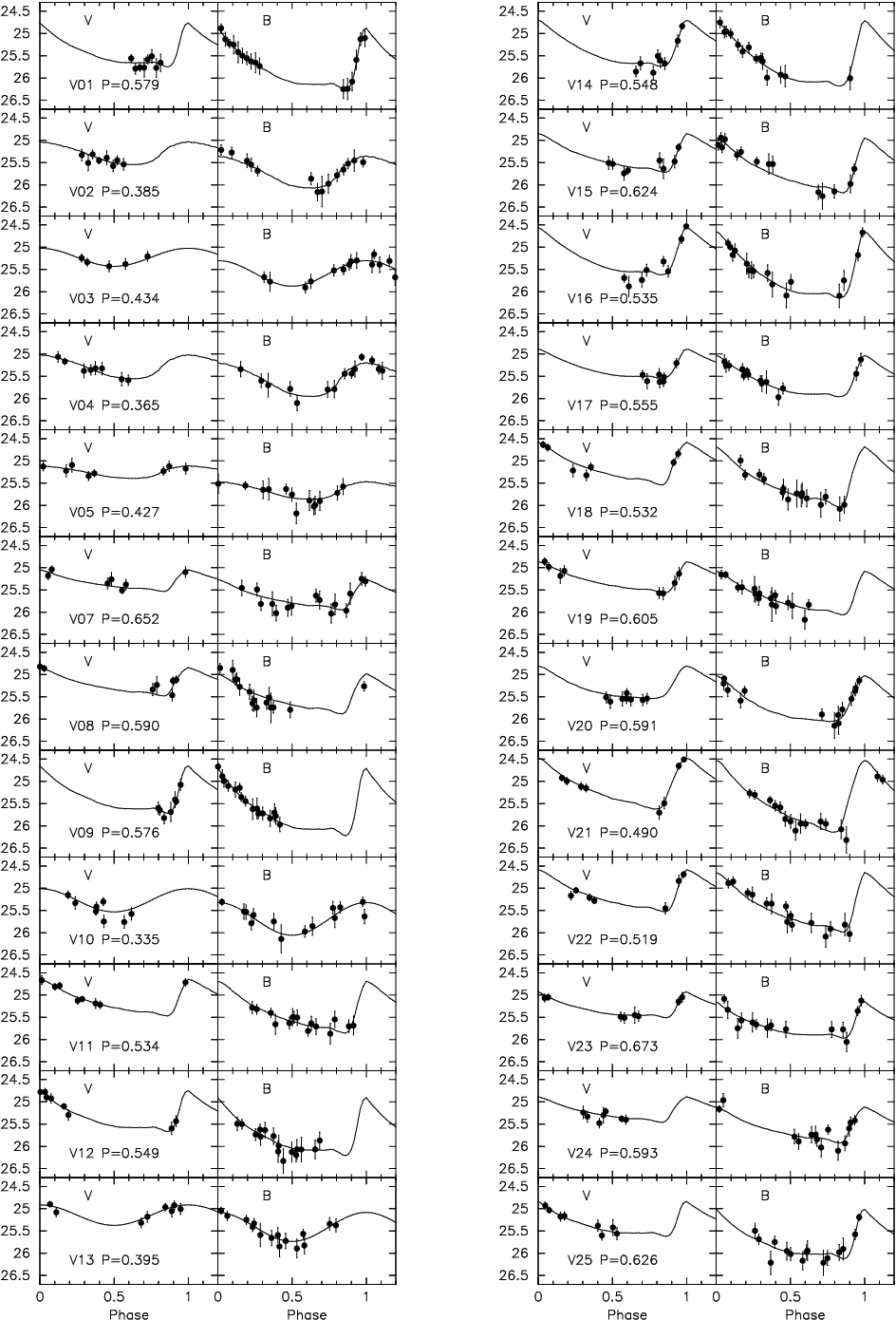,height=7.75in,width=5.5in}} 
  \caption{And~VI RR~Lyrae light curves with the observations shown as filled 
           circles and the fitted template displayed as a curve.} 
  \label{Fig04}
\end{figure*}

\begin{figure*}[t]
  \figurenum{4 cont}
  \centerline{\psfig{figure=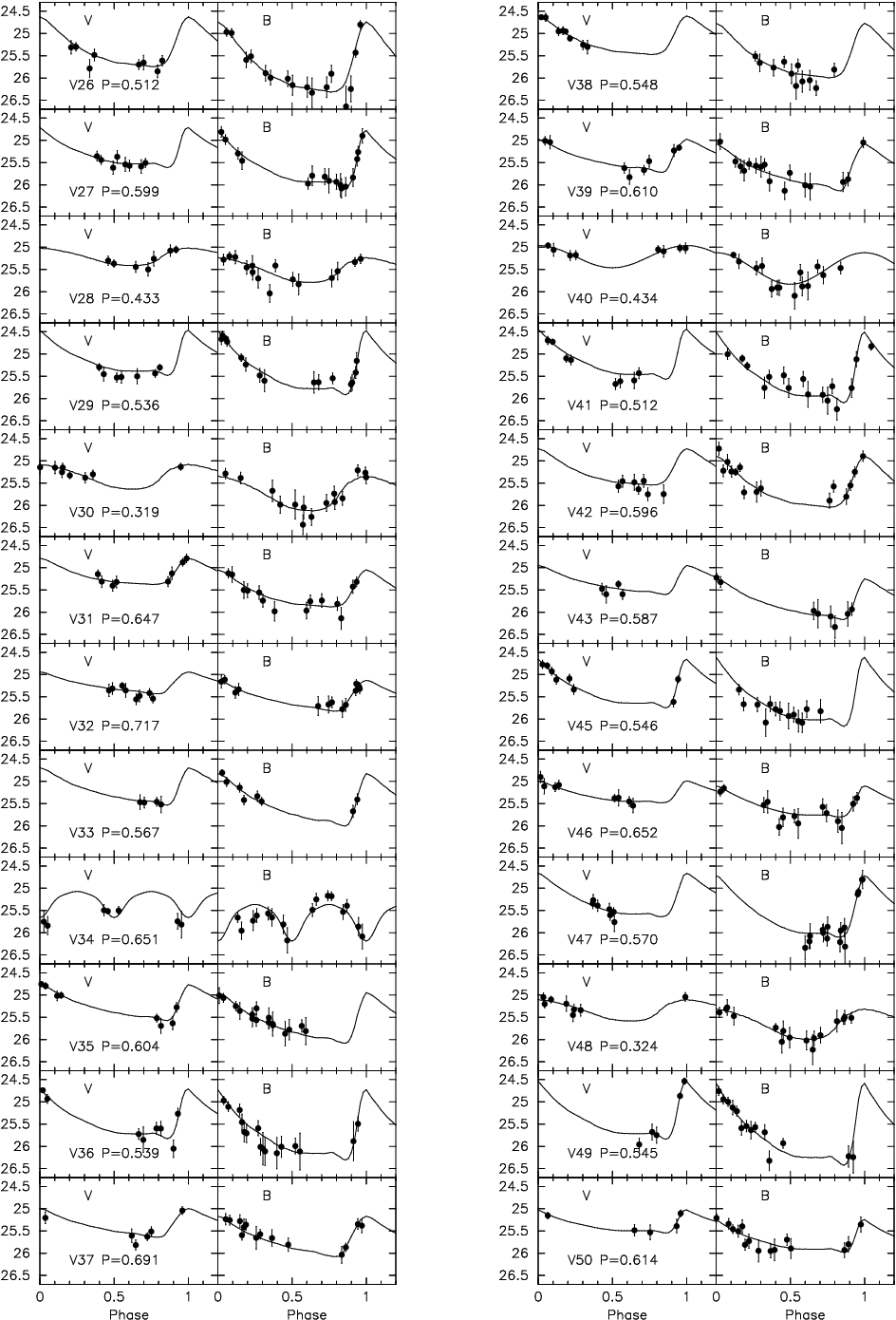,height=7.75in,width=5.5in}} 
  \caption{And~VI RR~Lyrae light curves with the observations shown as filled 
           circles and the fitted template displayed as a curve.} 
  \label{Fig04}
\end{figure*}

\begin{figure*}[t]
  \figurenum{4 cont}
  \centerline{\psfig{figure=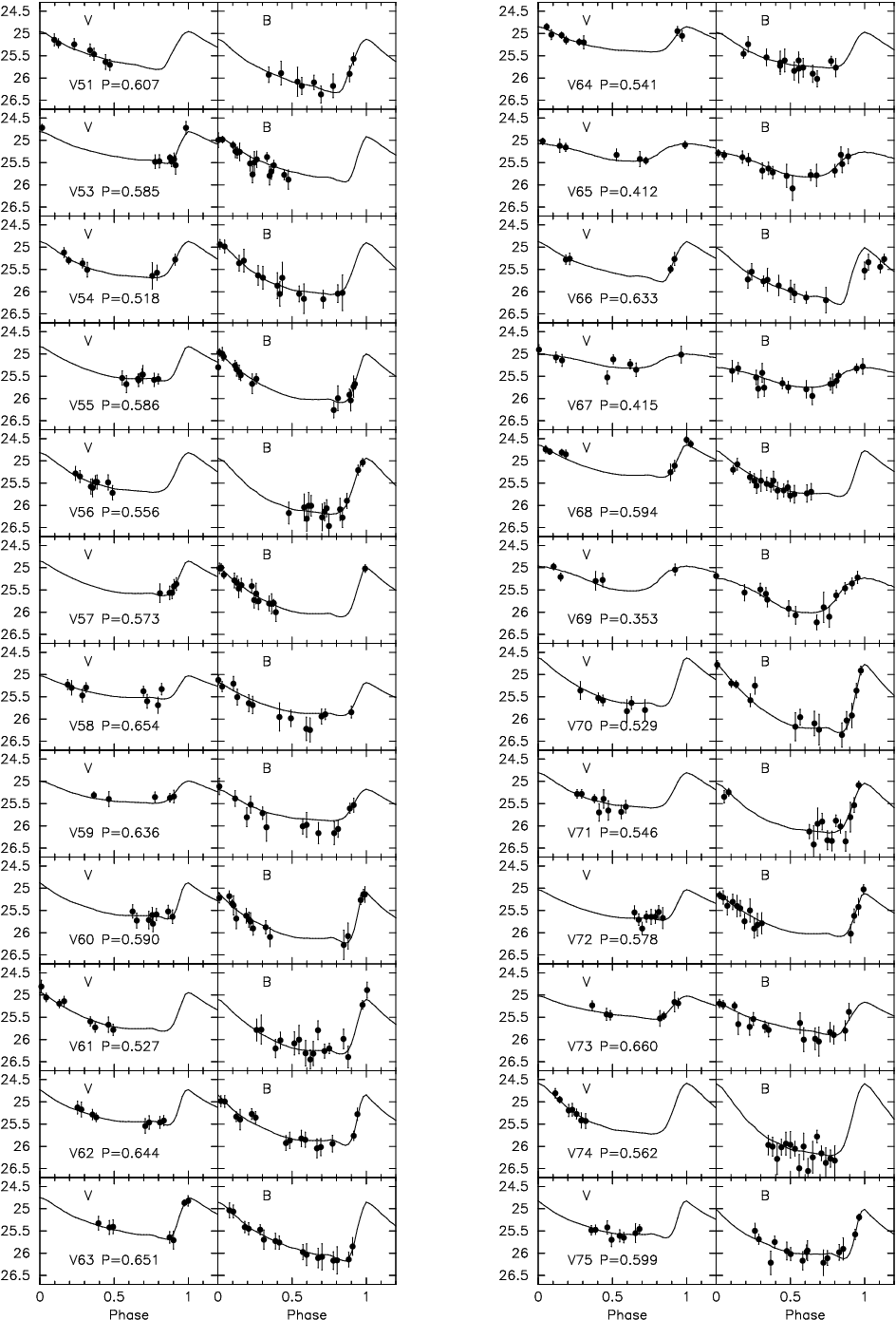,height=7.75in,width=5.5in}} 
  \caption{And~VI RR~Lyrae light curves with the observations shown as filled 
           circles and the fitted template displayed as a curve.} 
  \label{Fig04}
\end{figure*}

 \begin{figure*}[t]
  \figurenum{4 cont}
  \centerline{\psfig{figure=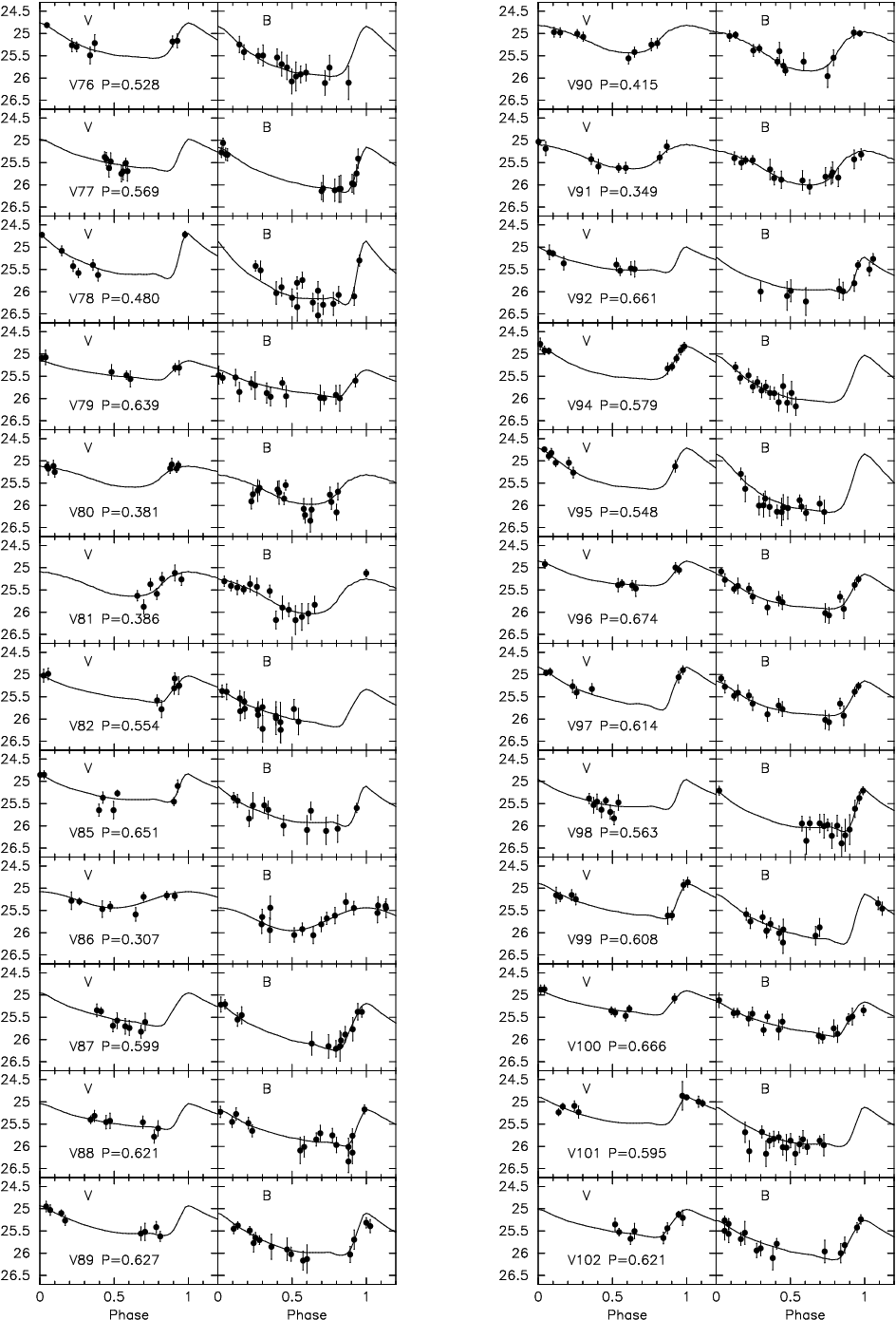,height=7.75in,width=5.5in}} 
  \caption{And~VI RR~Lyrae light curves with the observations shown as filled 
           circles and the fitted template displayed as a curve.} 
  \label{Fig04}
\end{figure*}

 \begin{figure*}[t]
  \figurenum{4 cont}
  \centerline{\psfig{figure=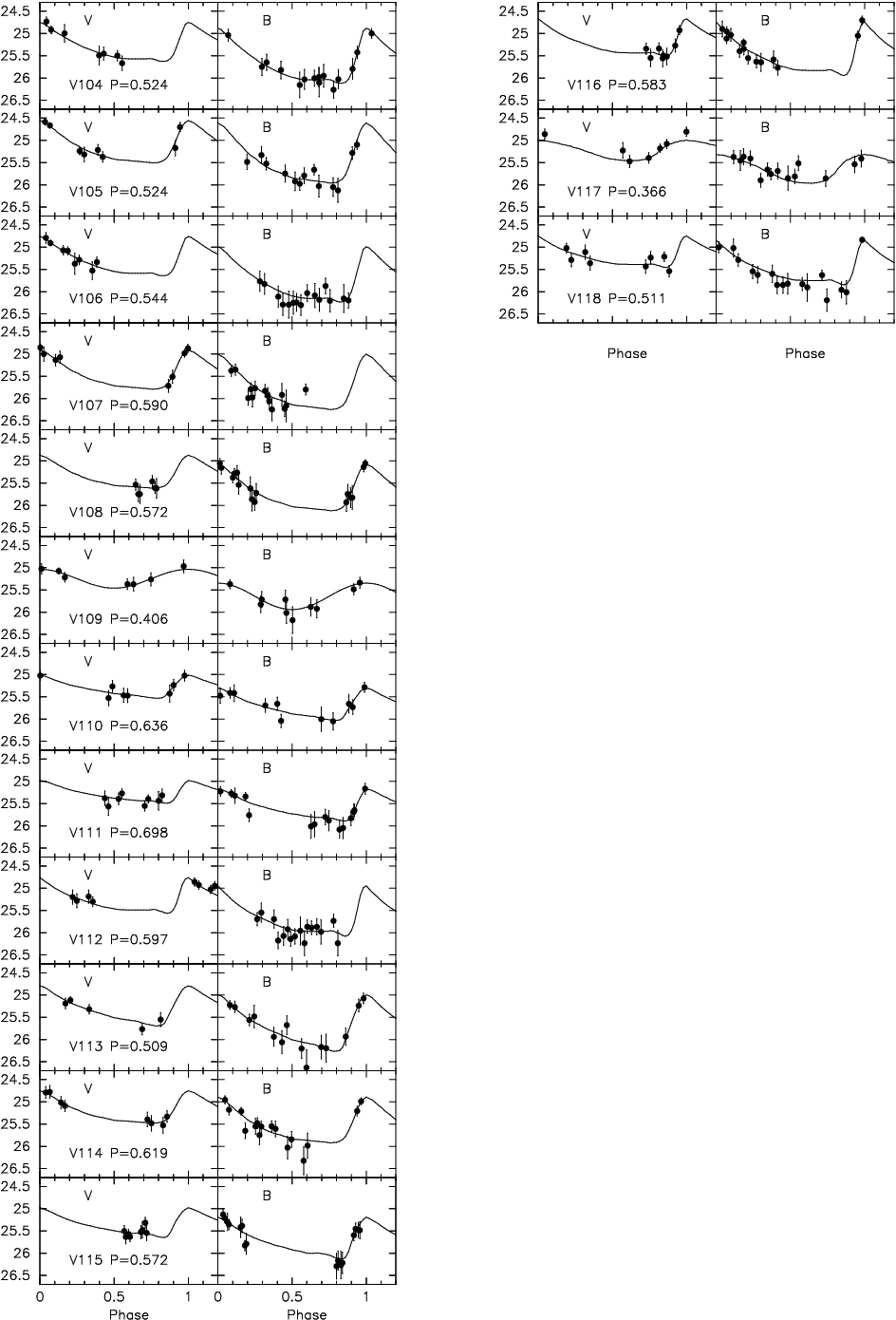,height=7.75in,width=5.5in}} 
  \caption{And~VI RR~Lyrae light curves with the observations shown as filled 
           circles and the fitted template displayed as a curve.} 
  \label{Fig04}
\end{figure*}

\section{And VI Anomalous Cepheids} 

ACs are typically 0.5 - 1.5 mag brighter than the RRL in a system.  This 
difference indicates that ACs have masses in the range of 1 - 2 $M_\odot$ 
(Norris \& Zinn 1975; Zinn \& Searle 1976; Zinn \& King 1982; Smith \& Stryker 
1986; Bono et al.\ 1997b).  In order for such stars to evolve into the 
instability strip, they are required to have low metallicities, approximately 
${\rm [Fe/H]} < -1.3$ (Demarque \& Hirshfeld 1975).  
Given that masses in excess of the turnoff mass of old globular clusters 
($M>0.8~M_{\odot}$) are required, the two leading hypotheses for the origin 
of ACs are that they are either stars from an intermediate age population, 
age less than 5 Gyr (e.g., Demarque \& Hirshfeld 1975; Norris \& Zinn 
1975), or stars of increased mass due to mass 
transfer in a binary system of older stars (Renzini, Mengel, \& Sweigart 
1977).  Both scenarios are plausible.  
It may be the case that the origin of the ACs is tied to the system in 
which they originate.  While there seems to be no clear way of selecting 
between the two origin scenarios, the most likely explanation for the 
ACs found in globular clusters, such as NGC~5466 (Zinn \& Dahn 1976) and 
possibly $\omega$ Centauri (Wallerstein \& Cox 1984), is the mass transfer 
scenario.  For dSphs, both scenarios may be effective.  

While ACs are almost nonexistent in globular clusters, they are 
common in dSphs.  Every dSph surveyed for variable stars 
has been shown to include at least one AC.\@  Thus it is not unexpected that 
And~VI would contain at least one AC, especially given the mean metallicity 
of the dSph ($\langle {\rm [Fe/H]} \rangle = -1.58$, AJD99). 

Six of the 118 variables found in this survey are likely ACs.  
Their photometric properties are listed in Table~1 and their 
light curves are shown in Figure~3.  The first and fourth columns of Table~1 
list the star's identification along with its period.  The 
intensity-weighted $\langle V \rangle$ and $\langle B \rangle$ magnitudes 
are given in the fifth and sixth columns.  The magnitude-weighted \bv 
color, $(\bv)_{\rm mag}$, is listed in the seventh column.  Columns eight and 
nine list the 
$V$ and $B$ amplitudes for the variables.  The other columns will be discussed 
subsequently, but the classifications are in column twelve.  We searched the 
literature for ACs with $(\bv)$ colors and found that they lie between the 
typical red and blue edges of the RRL instability strip.  As can be seen in 
Figure~1, the And~VI candidate ACs are consistent with the expected position 
for ACs in the CMD.\@  For two ACs we were not able to make full use of the 
data.  V44 was very near the edge of the WFC2 CCD in the second pointing, 
and so only photometry from the first pointing is available.  For V83, the 
star fell near or on a bad column in the first pointing.  This resulted in 
the loss of all the F450W data for this star at this pointing, though the 
F555W photometry was unaffected.  Similarly, for the second pointing, we 
were able to only use two F555W and two F450W measurements for this star.  
As a result, the magnitudes and colors for V44 and V83 are less well 
determined than for the other variables.  However, their periods and absolute 
magnitudes are consistent with those for the other ACs.

One way to classify the pulsation modes of variable stars is on the basis 
of the shape of their light curves.  First-overtone mode stars typically 
have a more sinusoidal shape to their light curves, while fundamental mode 
stars have more asymmetric light curves.  This is clearly seen in RRL and 
other Cepheid stars, but the case is not so clear for ACs.  While some 
first-overtone mode ACs may have more sinusoidal light curves, others look 
``less asymmetric" than a light curve for a fundamental mode star.  An example 
of this would be the difference between a RRa star, which has a sharp rise to 
maximum light, and a RRb star, which has a gentle rise to maximum light 
(Bailey 1902), though both are fundamental mode pulsators.  On this basis, 
we can attempt to classify the ACs in And VI.\@  
And VI ACs pulsating in the fundamental mode are V06 and V84.  First-overtone 
mode stars are V52 and V93.  We are unable to definitively classify 
V44 and V83 due to the small number of points and difficulties with the 
photometry as stated above.  On the other hand, the small number of points 
renders these classifications preliminary.  For example, from the absolute 
magnitude versus the period of the ACs, V93 may well be pulsating in the 
fundamental mode (see \S4.1).

\subsection{Anomalous Cepheid Absolute Magnitudes} 

In order to compare the properties of the And~VI ACs with those for other 
known ACs, we have converted the available photometry to absolute magnitudes.  
AJD99 derive an And~VI distance modulus $(m-M)_0=24.45$ from the $I$ magnitude 
of the tip of the red giant branch.  We have adopted this value together with 
their listed And~VI reddening, $E(\bv)=0.06$~mag.  

For the ACs in Galactic dSphs, we have adopted reddenings 
from Mateo (1998), while for the AC in the globular cluster NGC~5466 we 
have used the reddening from Harris (1996).  In order to have all distance moduli 
on the same system, we have searched the literature for distance estimates 
derived from the tip of the red giant branch.  For the cases where these 
estimates were not available, we used the mean magnitude of the RRL along 
with the absolute magnitude based on the Lee, Demarque, \& Zinn distance 
scale (Eq.\ 7 of Lee, Demarque, \& Zinn 1990) to calculate the distance modulus.  
The tip of the red giant branch method is based upon this distance scale.  
We have also adopted total-to-selective 
extinction ratios $R_V=3.1$ and $R_B=4.1$, respectively.  The results are given 
in Table~4.  Here the first column identifies the system while the second 
and third give the associated distance modulus and reddening.  The ID for 
each AC and the mode of pulsation (F$=$fundamental mode; H$=$first-overtone 
mode) based on the position in the absolute magnitude versus the logarithm 
of the period diagram (see below) are given in columns 4 and 5, respectively, 
along with the period (column 6).  Note that the IDs given for the ACs taken 
from Bersier \& Wood (2002) only list the last three digits of the full IDs 
given in that paper.  The mean apparent magnitudes and respective 
absolute magnitudes are listed in columns 7 through 10.  Columns 11 and 12 
list the $V$ and $B$ amplitudes for the ACs.  In each case we have reviewed 
the original photometry to check the periods and magnitudes.  Cases where we 
have made revisions to previous adopted values are indicated by an asterisk.  
We have not attempted to revise the Leo~I data because of its generally poorer 
quality.

\begin{figure*}[t]
  \centerline{\psfig{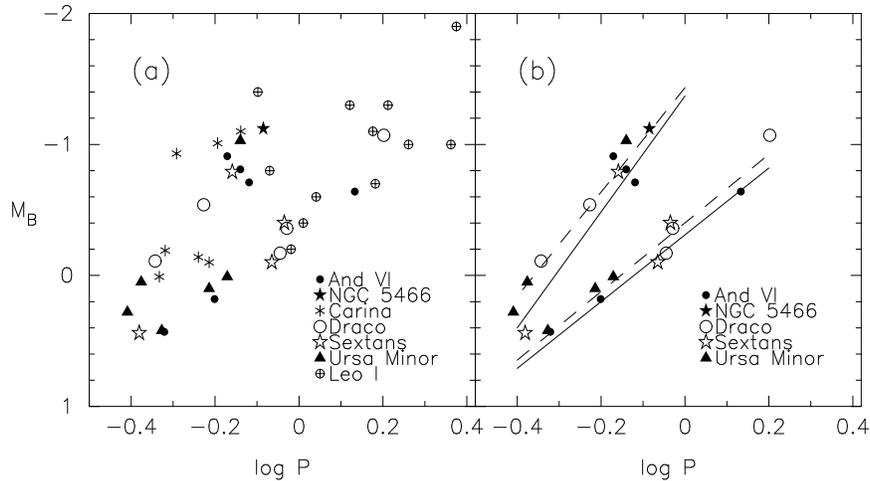}}
  \caption{$M_B$ versus period for anomalous Cepheids.  Figure (a) includes 
           all available anomalous Cepheids and (b) leaves out the Carina and 
           Leo~I data.  The solid lines in (b) are from Eqs.\ 13 and 14 from 
           Bono et al.\ (1997b), while the dashed lines are from Eqs.\ 2 and 
           3 from this paper.} 
  \label{Fig05}
\end{figure*}

Figure~5 shows the $B$ absolute magnitudes for the ACs as a function of their 
period.  In Figure~5a, which plots all available ACs, scatter 
is present in the plot and the difference between the two pulsational modes 
is not clear.  Since there is significant scatter in the photographic light 
curves of the Carina and Leo~I ACs, we have removed these ACs in Figure~5b.  
The difference between the pulsation modes is more evident in this plot.  
It is clear that the slopes for the two pulsation 
modes are not parallel.  This contradicts the suggestion of Nemec, Nemec, \& 
Lutz (1994) (cf.\ Figure~5), although the idea of non-parallel lines is in 
better agreement with the original analysis of Nemec, Wehlau, \& 
Mendes~de~Oliveira (1988).  The difference in slopes was also noted by Bono et 
al.\ (1997b) in their analysis of ACs based on the Nemec, Nemec, \& Lutz data.  
Using Eqs.\ 13 and 14 of Bono et al., we 
have plotted these lines in Figure~5b for the different pulsational modes.  
The lines appear to be slightly fainter than the data.  We have attempted to 
fit lines to the data in Figure~5b.  Due to the fundamental and first-overtone 
ACs converging at short periods, there is some uncertainty in the classification 
of those variables.  For the purposes of fitting the lines, we have left out 
V9 in Sextans since it is unclear whether it is pulsating in the fundamental 
or first-overtone mode.  The equations for the resulting lines are: 

\begin{equation}
M_{B,{\rm F}} = -2.62(\pm0.18)\log\,P - 0.40(\pm0.04) 
\end{equation} 
\begin{equation} 
M_{B,{\rm H}} = -3.99(\pm0.27)\log\,P - 1.43(\pm0.09) 
\end{equation} 

\noindent 
The slopes for our lines match well with what was found by Bono et al., but 
the zeropoint is different, perhaps as a result of different adopted 
distance scales.  

\begin{figure*}[t]
  \centerline{\psfig{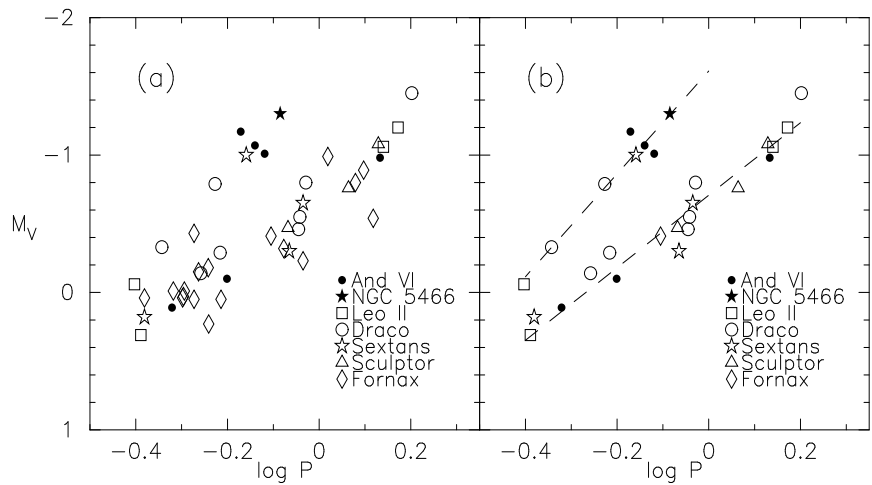}}
  \caption{$M_V$ versus period for anomalous Cepheids.  All data is included 
           in (a) and the Fornax data from Bersier \& Wood (2002) is removed 
           in (b).  The dashed lines in (b) are from Eqs.\ 4 and 5 from this 
           paper.} 
  \label{Fig06}
\end{figure*}

In Figure~6 we have plotted $M_{V}$ versus period.  Again, in Figure~6a 
we have included all available data.  Much of the scatter is due to the 
Fornax ACs from Bersier \& Wood (2002).  Because of the high scatter in their 
light curves, we have left out these variables in Figure~6b.  As for the $B$ 
data, we have fit lines to the data in Figure~6b.  We again omit from the fit 
V9 in Sextans, and also V1 in Leo~II for similar reasons which will be 
discussed in \S4.2. 

\begin{equation}
M_{V,{\rm F}} = -2.64(\pm0.17)\log\,P - 0.71(\pm0.03) 
\end{equation} 
\begin{equation} 
M_{V,{\rm H}} = -3.74(\pm0.20)\log\,P - 1.61(\pm0.07) 
\end{equation} 

\noindent 
The slopes for these lines match well to those found for the $B$ AC data.  
We are not aware of any {\it a priori} reason why this should be true.

\subsection{Anomalous Cepheid Period-Amplitude Diagram} 

In Figure~7 we plot the amplitudes versus log period for our sample of ACs.  
The Carina and Leo~I ACs have been excluded from Figure~7a due to the large 
scatter in their photometry, while the Fornax ACs from Bersier \& Wood (2002) 
have been left out of Figure~7b for similar reasons.  The pulsation modes are 
assigned according to each variable's positions in the $M-\log\,P$ diagrams.  
In Figure~7 there is no clear trend in $V$ or $B$ amplitude with pulsation 
mode.  This is similar to what was found by Nemec, Nemec, \& Lutz (1994, their 
Fig.~10).  The same is not true for RRL where there are clear distinctions 
between pulsation modes in the period-amplitude diagram.  For RRL, there is a 
clear break in period between fundamental mode, RRab, and first-overtone mode, 
RRc, stars with the longest period RRc star having a shorter period than the 
shortest period RRab star in a given system.  This is due to the ionization 
zone being located deeper within the envelope for the bluer variables.  A 
similar effect is seen in the classical Cepheids (e.g., Smith et al. 1992) 
where the first-overtone Cepheids tend to have lower amplitudes coupled with 
their shorter periods.

Bono et al.\ (1997b) used theoretical models to analyze the properties of ACs.  
They argued that the predicted location of the zero-age horizontal branch in 
the amplitude-$\log\,P$ diagram provided a transition between the 
fundamental mode pulsators and the first-overtone mode pulsators in the 
period-amplitude diagram.  We have plotted this line in Figure~7b.  Bono et 
al.\ stated that all ACs with period shorter than this line are pulsating in 
the first-overtone and those with longer periods are pulsating in the 
fundamental mode.  

Taking a closer look at the fundamental mode ACs to the 
left of the Bono et al.\ line, the likely reasons for their being 
to the left of the line are either they have the wrong classification or their 
periods and/or amplitudes are incorrect.  V6 from And~VI is close enough to 
the line that we leave it out of further discussion.  V93 from And~VI is 
another one of the fundamental mode ACs to the left of the line.  Due to the 
fundamental mode and first-overtone mode lines nearly intersecting as seen in 
Figures~5 and 6, it is possible that V93 is actually a first-overtone pulsator 
although it doesn't seem likely.  A better case may be made for V9 in Sextans.  
Still, there are a few other ACs (V1 and V56 of Ursa Minor and V055 and V208 
of Draco) that are found 
among the first-overtone pulsators that appear to be pulsating in the 
fundamental mode according to their position in the absolute magnitude versus 
$\log\,P$ plots.  In any case, more and better observations 
would likely help in better defining the period-amplitude diagrams.

\section{And~VI RR~Lyrae} 

The presence of RRL within a system indicates the existence of an old 
population ($t>10$~Gyr).  For And~VI we have detected 111 RRL 
with 91 pulsating in the fundamental mode (RRab) and 20 pulsating in 
the first-overtone mode (RRc).  There is one RRL (V103) which we were unable 
to place on the $B$,$V$ system due to a large gap in its light curve.  As a 
result, its magnitude and amplitude were not used in the following results.  
Although our search for variables was 
extensive, the number of RRc stars is more than likely greater than what 
we found.  The scatter in the photometry along with the low amplitude of the 
RRc star makes it difficult to detect this type of variable.  A histogram of 
the RRL periods in And~VI is compared to other dSphs in Figure~8, with the 
dSphs increasing in mean metallicity from the top, Ursa Minor, to the bottom, 
Fornax.  The sources for the data are the same as those listed in Table~4.  
There is a clear trend in the populations of the RRab stars with the more 
metal-poor dSphs tending to have longer period RRab stars than the more 
metal-rich dSphs.  There appears to be no obvious trend with the RRc stars.  
This may be due, in part, to the difficulty in detecting the smaller amplitude 
RRc stars, as we noted in this survey. 

\begin{figure*}[t]
  \centerline{\psfig{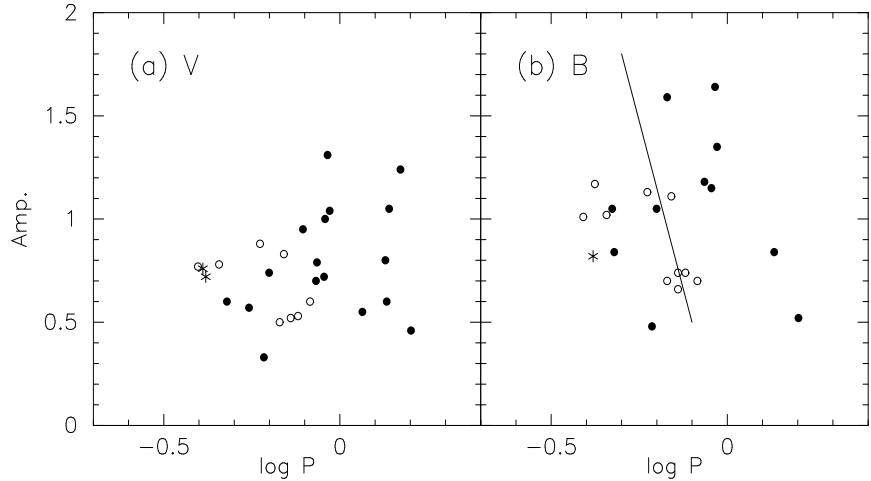}}
  \caption{Period-amplitude diagram for anomalous Cepheids.  First-overtone 
           stars are denoted by open circles and fundamental mode stars are 
           denoted by filled circles.  Asterisks denote ACs with uncertain 
           classification as noted in Table~4.  Only anomalous Cepheids with 
           good photometry are included in the $V$ (a) and $B$ (b) plots.  
           The line in (b), taken from Figure~7 of Bono et al.\ (1997b), 
           represents the predicted location of the zero-age horizontal branch 
           fundamental mode pulsators, and is meant to divide fundamental mode 
           from first-overtone pulsators.} 
  \label{Fig07}
\end{figure*}

\begin{figure*} 
  \centerline{\psfig{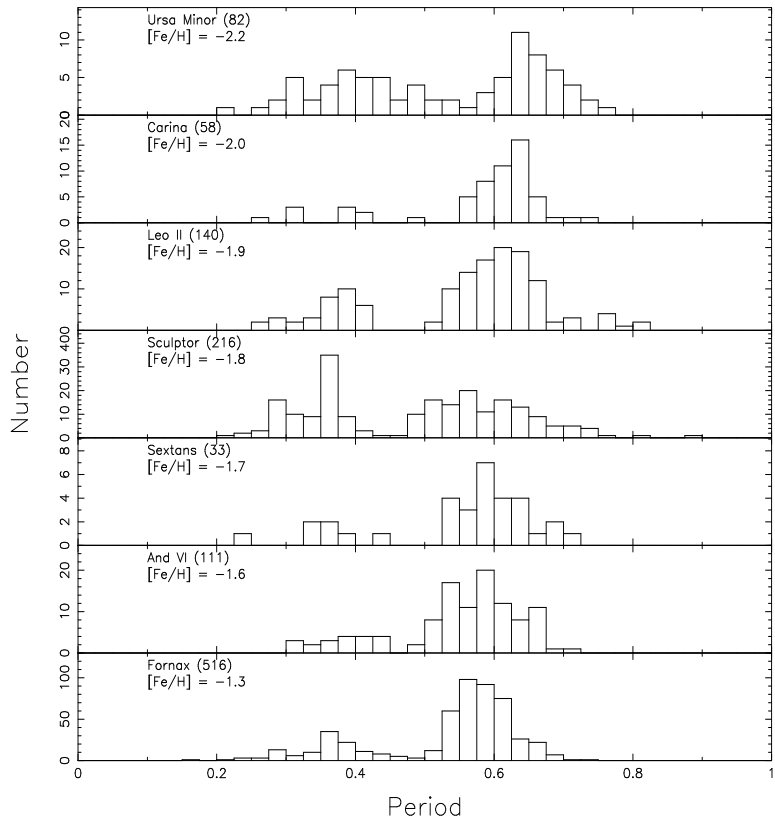}} 
  \caption{Period distribution plots for the RR~Lyrae in dwarf spheroidal 
           galaxies.  The [Fe/H] values were taken from Mateo (1998).}
  \label{Fig08}
\end{figure*}

From the sample of RRL in And~VI we find the mean magnitude to be 
$\langle V \rangle = 25.29\pm0.03$, where the uncertainty is the aperture 
correction uncertainty, the photometry zeropoint uncertainty, and the 
spline-fitting uncertainty added in quadrature to the 
standard error of the mean.  In order to convert this RRL magnitude to a 
distance on the same scale as the tip of the red giant branch distance of 
AJD99, we calculated $M_V$ of the RRL from Lee, Demarque, \& Zinn (1990), 

\begin{equation} 
M_{V,{\rm RR}} = 0.17{\rm [Fe/H]} + 0.82 {\rm .} 
\end{equation} 

\noindent 
Adopting $\langle {\rm [Fe/H]} \rangle=-1.58\pm0.20$ and $E(\bv)=0.06\pm0.01$ 
from AJD99, we find $M_{V,{\rm RR}} = 0.55$ and 
$A_V = 3.1E(\bv) = 0.19\pm0.03$.  The resulting distance is $815\pm25$~kpc.  
This matches up well with the tip of the red giant branch distance estimate by 
AJD99 of $775\pm35$~kpc.

\subsection{RR~Lyrae Period-Amplitude Diagram}

The period-amplitude diagram provides an important diagnostic tool when 
investigating the properties of a system as the period and the amplitude 
of a variable are independent of quantities such as distance and reddening.  
It is generally thought that the position of a RRL in the period-amplitude 
diagram is dependent on the metallicity of the star (Sandage 1993b).  More 
recently, Clement et al.\ (2001) in a study of the properties of the RRL in 
Galactic globular clusters suggested that while this may be true for the 
more metal-rich Oosterhoff type I clusters, the same may not be true for the 
more metal--poor Oosterhoff type~II clusters.

In Figure~9, we plot the RRL found in And~VI in a period-amplitude ($A_B$) 
diagram.  The RRL fall in the expected positions with the RRc stars at shorter 
periods and lower amplitudes as compared to the longer period RRab stars.  
The RRc stars appear to fall into a parabolic shape, an effect predicted by 
Bono et al.\ (1997a). The width seen in the RRab stars is similar to what is 
seen in other dSphs (e.g., Leo~II: Siegel \& Majewski 2000; Sculptor: Kaluzny 
et al.\ 1995).  This is not unexpected since dSphs are known to have a spread 
in their metallicity (see Mateo 1998 and references therein).  It is uncertain 
how much the And~VI spread in metallicity ($\sigma({\rm [Fe/H]})\approx0.3$, 
AJD99) may be the cause of the observed spread of the RRab stars in 
the period-amplitude diagram and how much may be due to other effects such as 
age or evolutionary effects. 

\begin{figure*}[t]
  \centerline{\psfig{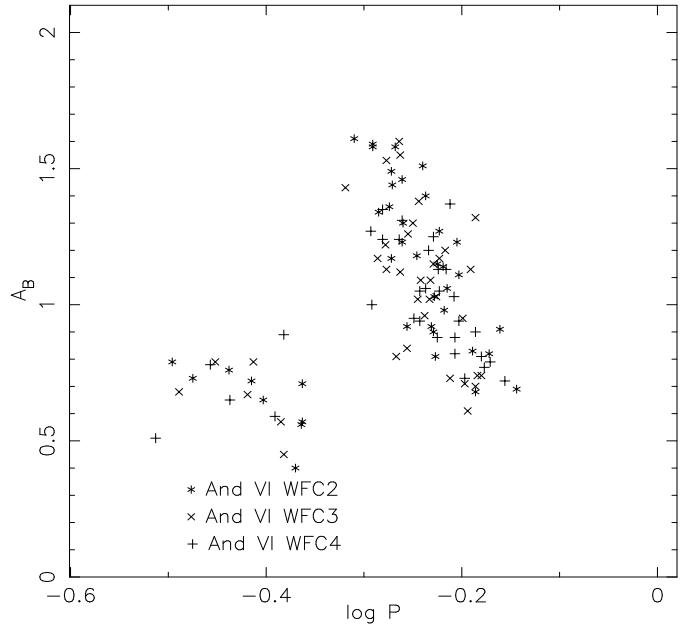}}
  \caption{Period-amplitude diagram for the RR~Lyrae in And~VI.\@  The 
           amplitudes shown are for the $B$ filter.} 
  \label{Fig09} 
\end{figure*}

\begin{figure*}[t]
  \centerline{\psfig{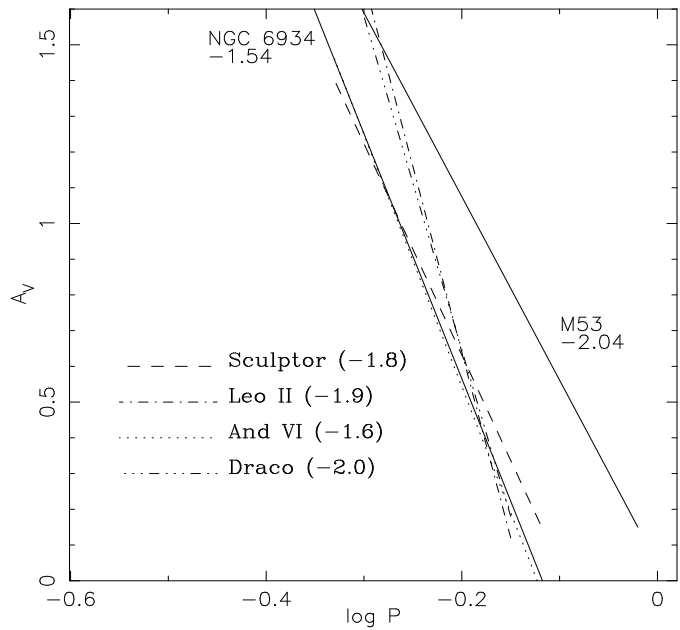}}
  \caption{Period-amplitude diagram showing the linear regression fits to the 
           RR~Lyrae in dwarf spheroidal galaxies.  Also shown are the linear 
           regression fits to NGC~6934 and M53.} 
  \label{Fig10} 
\end{figure*}

To allow comparison to other dSphs, we fit the RRab stars in the 
period-amplitude diagram by linear regression fits using the second equation 
from Table~1 of Isobe et al.\ (1990).  This version of ``least squares," which 
places the dependence on the x-variable against the independent y-variable and 
accounts for uncertainties in both variables, gives the 
best fit visually to the data.  Figure~10 plots the fits for And~VI, Leo~II 
(Siegel \& Majewski 2000), Draco (Kinemuchi et al.\ 2002), and Sculptor 
(Kaluzny et al.\ 1995).  For reference we also plotted the fits to two 
Oosterhoff clusters:  NGC~6934 (Oosterhoff type~I; Kaluzny, Olech, \& Stanek 
2001) and M53 (Oosterhoff type~II; Kopacki 2000).  These 
two clusters were chosen due to their well-defined data.  It should be noted 
that these globular cluster lines are similar to the Oosterhoff lines defined 
by Clement (2000 and private communications).  Table~5 lists the equations for 
each system.  The fit for the RRab stars in And~VI coincides with the fit for 
NGC~6934.  This agrees with the idea that the position of RRab stars in a 
period-amplitude diagram is dependent on the metallicity since the 
metallicities of NGC~6934 (${\rm [Fe/H]}=-1.54$) and And~VI 
($\langle {\rm [Fe/H]} \rangle =-1.58$) are the same.

The fit of the Sculptor data, although slightly shifted toward longer periods, 
is found to be similar to the line for the more metal-rich NGC~6934.  This is 
a bit surprising given the metallicity of Sculptor 
($\langle {\rm [Fe/H]} \rangle =-1.8$).  With [Fe/H] values of -2.0 and -1.9 
for Draco and Leo~II, respectively (Mateo 1998), one would expect the RRab 
stars to fall near the M53 (${\rm [Fe/H]}=-2.04$) line in Figure~10.  Instead 
the fits to the data fall somewhere between NGC~6934 and M53 lines.  The slopes 
of the lines for Leo~II 
and Draco, while similar to each other are different from the slopes of 
And~VI and Sculptor which also are similar to each other.  Although the exact 
reason for the differences seen in the slopes and locations of the fits are 
uncertain, it is an effect that is seen in other 
dSphs, where the mean periods of the RRab stars in dSph systems do not follow 
the shift in period with abundance seen in Galactic GCs.  We discuss this 
in the next section.

\subsection{Oosterhoff Classification}

\begin{figure*}[t]
  \centerline{\psfig{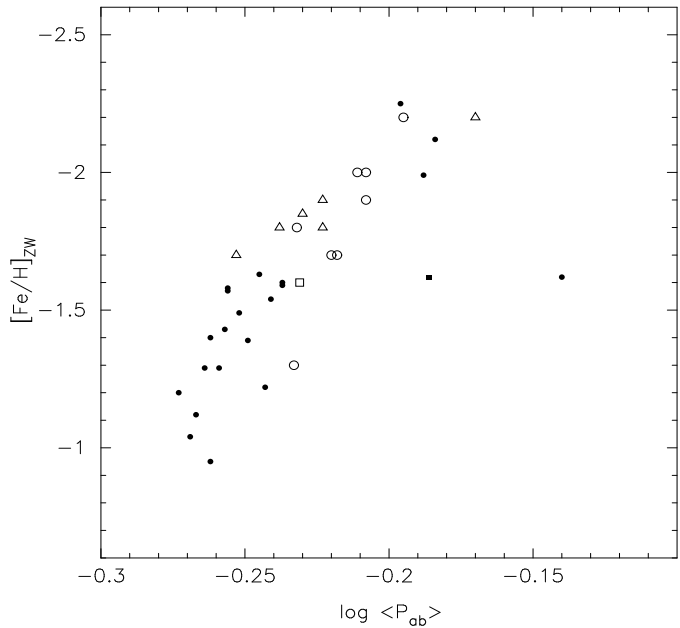}}
  \caption{The mean period for the RRab stars versus the metallicity of the 
           parent system.  Galactic globular clusters are shown as filled 
           circles, along with $\omega$~Centauri (filled square).  Large 
           Magellanic Cloud globular clusters are shown as open triangles.  
           The Galactic dwarf spheroidals are indicated by open circles.  
           And~VI is shown as an open square.} 
  \label{Fig11} 
\end{figure*}

The RRab stars in metal-poor globular clusters have longer mean periods 
($\langle P_{ab} \rangle = 0.64$d) and higher ratios of RRc stars compared to 
the total number of RRL ($N_c/N_{RR}=0.44$) than do RRab stars in metal-rich 
clusters (Oosterhoff 1939) where $\langle P_{ab} \rangle = 0.55$d and 
$N_c/N_{RR}=0.17$ (see Smith 1995 and references therein).  Further, Galactic 
globular clusters having $-2.0<{\rm [Fe/H]}<-1.7$ contain few RRL because 
their horizontal branches lack stars in the vicinity of the instability 
strip.  Although it is clear that metallicity is the first parameter governing 
the morphology of the horizontal branch and therefore the RRL content within 
the instability strip, secondary parameters (e.g., age, mass loss 
along the red giant branch, and rotation) also affect the horizontal 
branch morphology and thus confuse the general Oosterhoff trend.

RRL in the Galactic dSphs fall along a continuum within the Oosterhoff 
gap (van Agt 1973; Zinn 1978, 1985; Kaluzny et al.\ 1995; Mateo et al.\ 1995).  
A number of the Large Magellanic Cloud (LMC) 
globular clusters also fall within the Oosterhoff gap.  In Table~6 we list 
the properties of the dSphs, including And~VI, and the LMC 
globular clusters with at least 15 RRab stars whose light curves have 
minimum scatter and gaps in the data.  For comparison, the mean properties 
for the Oosterhoff groups are also listed.  The first two columns identify 
the various systems.  The metallicities for the dSphs in the third column 
were taken from Mateo (1998; except for Leo~I which is from Gallart et al.\ 
1999), while the metallicities for the LMC GCs were 
taken from the literature.  Columns 4 and 5 give the mean periods for the 
RRab and RRc stars and columns 6--8 give the number of RRab and RRc stars, 
along with the ratio of RRc stars to the total number of RRL.  To better 
illustrate how these systems fill in the ``intermediate" metallicities we have 
also plotted the logarithm of the mean RRab periods versus the metallicity of 
the parent system in Figure~11.  We have left out the Sagittarius dSph since 
the precise metallicity related to the RRL is uncertain.  Also plotted in the 
figure are Galactic globular clusters with at least 15 RRab stars (Clement 
et al.\ 2001 and references therein).  Excluding $\omega$~Centauri, which has 
a spread in metallicity (e.g., Freeman \& Rodgers 1975), and M2, which shows 
the second parameter effect (e.g., Lee \& Carney 1999), the Oosterhoff 
dichotomy is clearly seen in the Galactic globular clusters as noted by 
Oosterhoff (1939).  As discussed above, the properties of And~VI are consistent 
with the Galactic dSphs.  The one Galactic dSph that is somewhat offset from 
the other dSphs in Figure~11 is Fornax.  Given the large age and abundance 
range in Fornax, it is possible that the mean [Fe/H] of the RRL is lower than 
that of the galaxy as a whole (see Bersier \& Wood 2002). This effect would act 
to displace Fornax toward the other dSphs in Figure~11.

While studying M15 and M3, Sandage, Katem, and Sandage (1981) noticed a 
period shift between the RRL in these clusters in the period-amplitude diagram 
that correlated with metallicity (Sandage 1981;1982a,b).  Metal-poor clusters 
tend to have longer periods for a given amplitude when compared to metal-rich 
clusters such that, 

\begin{equation} 
\Delta\log\,P = -(0.129A_B + 0.088 + \log\,P)
\end{equation}

\noindent 
relative to M3 (Sandage 1982a,b).  From Eq.~7, Oosterhoff type I clusters have 
$\langle\Delta\log\,P\rangle \ge -0.01$, while Oosterhoff type II clusters 
have $\langle\Delta\log\,P\rangle \le -0.05$ and field RRL tend to avoid 
a range $-0.05 \le \Delta\log\,P \le -0.01$ (Suntzeff et al. 1991).  

RRL in dSphs, though, are found in the 
$-0.05 \le \Delta\log\,P \le -0.01$ range (Sextans:  Mateo et al.\ 
1995; Leo~II:  Siegel \& Majewski 2000).  For And~VI we calculated the 
$\Delta\log\,P$ values for the RRab stars, listed in column ten of Table~1, 
and plot them against $\log\,P$ in Figure~12.  There are a number of stars 
in the region where Galactic RRab stars are absent.  This shows that the 
``intermediate" mean periods of the RRab stars in dSphs are not a result of a 
superposition of Oosterhoff type~I and II populations.  Still, it is 
interesting to note that $\langle\Delta\log\,P\rangle = 0.00$ for And~VI 
arguing that there is no period shift compared to the Oosterhoff type~I 
cluster, M3.  Again, this is not unexpected since the metallicity of M3 
(${\rm [Fe/H]} = -1.57$, Harris 1996) is the same as And~VI 
($\langle {\rm [Fe/H]} \rangle = -1.58$, AJD99).

\begin{figure*}[t]
  \centerline{\psfig{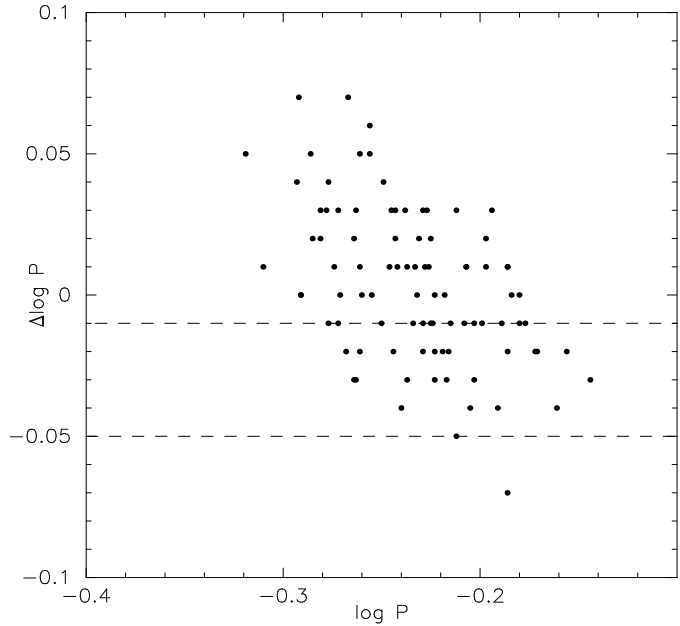}}
  \caption{The period shift versus period for the RRab stars in And~VI.\@  The 
           dashed lines represent the zone in which few RRab stars from  
           Galactic globular clusters are found.} 
  \label{Fig12} 
\end{figure*}

It is easy to see from Table~6 that the mean periods of the RRab stars in the 
``intermediate" systems primarily follow the metallicity of the parent system.  
Therefore it is not surprising to see that $\langle P_{ab} \rangle$ for 
And~VI is similar to the values found in Galactic globular clusters of 
similar [Fe/H].  This is consistent with our finding in \S5.1 that the 
individual RRab stars in And~VI fall along the line defined by the Oosterhoff 
type~I cluster, NGC~6934 (${\rm [Fe/H]} = -1.54$, $\langle P_{ab} \rangle = 
0.574$d, Kaluzny et al.\ 2001).  

The presence of systems within the gap between the Oosterhoff 
groups brings up questions regarding whether there is an Oosterhoff dichotomy 
as seen in Galactic globular clusters or an ``Oosterhoff continuum" as seen 
in dSphs (Renzini 1983; Castellani 1983).  The question is:  What is the 
origin of the difference between systems such as the dSphs and LMC globular 
clusters that have significant numbers of RRL and the Galactic globular 
clusters in the same abundance range which have little or no RRL?  Clearly, 
the presence of RRL in this metallicity range is due to the 
differences in the horizontal branch morphology between the systems.    
All dSphs with adequately deep CMDs, except Ursa~Minor, exhibit the second 
parameter effect.  The combined effects of metallicity spreads, in addition to 
extended periods of star formation (Mateo 1998), may explain the different 
horizontal branch morphology in the dSph systems.  Yet it is unlikely this can 
be the explanation for the LMC globular clusters which have no metallicity 
spread and have been shown to have the same ages as their Galactic 
counterparts (Olsen et al.\ 1998; Johnson et al.\ 1999).  Further observations 
of other systems within this metallicity range would help in clearing up this 
debate.  In any case, the idea of metallicity being the first parameter 
governing the horizontal branch morphology and $\langle P_{ab} \rangle$ is 
clear in Figure~11.

\subsection{Metallicity Estimates from RR~Lyrae}

A number of Galactic dSphs have been shown to exhibit a spread in their 
metallicities (see Mateo 1998 and references therein).  Consequently, the 
exact metallicities for the individual RRL are uncertain.  One method 
for determining the metallicity of the bulk of the RRL is through 
a relation determined by Alcock et al.\ (2000) relating the metallicity of 
a RRab star to its period and $V$-band amplitude.  The relation, 

\begin{equation}
{\rm [Fe/H]}_{\rm ZW} = -8.85(\log\,P_{ab} + 0.15A_V) - 2.60{\rm ,} 
\end{equation}

\noindent
was calibrated using the RRab stars in M3, M5, and M15, where ZW refers to the 
Zinn \& West (1984) scale.  Their calibration predicted the metallicity 
of the RRL in those systems with an accuracy of $\sigma_{\rm [Fe/H]} = 0.31$ 
per star.  Alcock et al.\ tested this formula on their 
sample of RRab stars from the LMC and found the resulting median metallicity 
to be in good agreement with previous results. 

We list in column eleven of Table~1 the resulting metallicities for the 
individual RRab stars using Eq.\ 8.  These results were plotted in the [Fe/H] 
distribution plot seen in Figure~13.  A gaussian fit to the data reveals a 
mean metallicity of ${\rm [Fe/H]} = -1.58$ with a fitting error of 0.01 and a 
standard deviation of 0.33.  This is exactly what was found by AJD99 
($\langle {\rm [Fe/H]} \rangle=-1.58\pm0.20$) using the red giant branch mean 
$V$--$I$ color.  Since the accuracy of the 
Alcock et al.\ equation is on the order of $\sim0.3$~dex, there is no evidence 
for any abundance dispersion in the RRab stars given $\sigma = 0.33$.  We 
also investigated the distribution of ${\rm [Fe/H]}$ values with distance 
from the center of And~VI and saw no convincing evidence for any metallicity 
gradient. 

\begin{figure*}[t]
  \centerline{\psfig{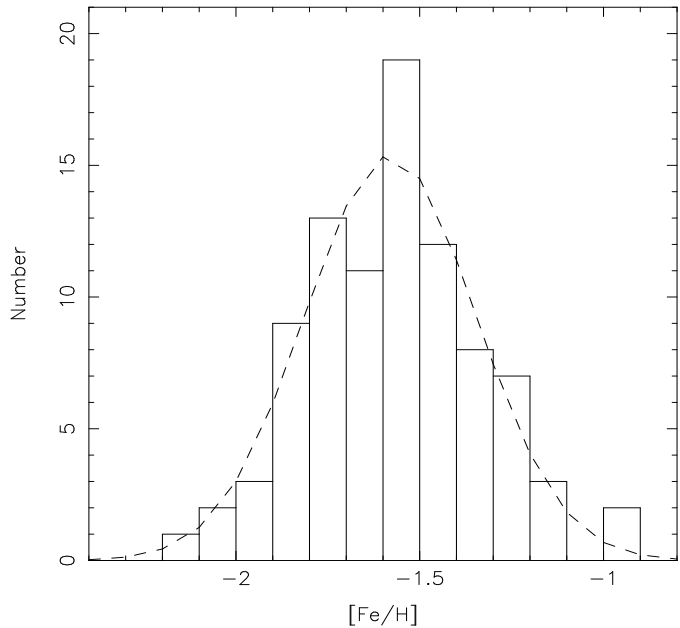}}
  \caption{[Fe/H] distribution plot for the RRab stars in And~VI.\@  The 
           individual [Fe/H] values were calculated using Eq.\ 2 in Alcock 
           et al.\ (2000).  The dashed line represents a gaussian fit to the 
           data.  The curve has a mean of ${\rm [Fe/H]}=-1.58$ and a standard 
           deviation of 0.33.} 
  \label{Fig13} 
\end{figure*}

Sandage (1993a) studied cluster and field RRL spanning a wide range 
of metallicities and related the average periods of RRL to their 
metallicities through the relations 

\begin{equation}
{\rm [Fe/H]}_{\rm ZW} = (-\log\,\langle\,P_{ab}\rangle - 0.389)/0.092 
\end{equation} 
\begin{equation} 
{\rm [Fe/H]}_{\rm ZW} = (-\log\,\langle\,P_{c}\rangle - 0.670)/0.119 
\end{equation}

\noindent
Siegel \& Majewski (2000) and Cseresnjes (2001) showed that, although these 
relations were derived from cluster and field RRL, the RRL in dSphs follow 
the same relations.  For And~VI, $\langle P_{ab} \rangle = 0.588\pm0.005$d and 
$\langle P_{c} \rangle = 0.382\pm0.005$d resulting in 
${\rm [Fe/H]}_{\rm ZW} = -1.72\pm0.04$ (internal error) for the RRab stars and 
${\rm [Fe/H]}_{\rm ZW} = -2.12\pm0.05$ (internal error) for the RRc stars.  The 
estimate derived from the RRab stars is slightly metal-poor when compared to 
the estimate derived from the Alcock et al.\ (2000) formula, but given the 
uncertainties in both methods, the values are not inconsistent with each 
other.  The metallicity estimate from the RRc stars is much more metal-poor 
than what was derived from the RRab stars.  This should be taken with some 
caution since as discussed in \S5, our search for RRc stars is more than likely 
incomplete.  As a result, the mean period for the RRc stars is probably 
inaccurate.

\section{Summary \& Conclusions} 

We have presented the light curves and photometric properties for 118 
variables found in the And~VI dwarf spheroidal galaxy.  The properties 
of the variable stars in And~VI were shown to be consistent with those found 
in other dSphs.  In particular, the 6 anomalous Cepheids found in 
And~VI have periods and absolute magnitudes similar to the anomalous Cepheids 
in other systems.  We redetermined the period-luminosity relations for the 
anomalous Cepheids and concur with the previous results of Nemec, Wehlau, \& 
Mendes de Oliveira (1988) and Bono et al.\ (1997b), that the lines 
representing the different pulsation modes are not parallel.  Unlike the 
situation for RR~Lyrae, we were not able to make 
a clear distinction between the fundamental and first-overtone mode anomalous 
Cepheids in a period-amplitude diagram.

From a sample of 110 RR~Lyrae, we found the mean $V$ magnitude to be 
$25.29\pm0.03$ resulting in a distance for And~VI of $815\pm25$~kpc on 
the Lee, Demarque, \& Zinn (1990) distance scale.  The mean period of the 
RRab stars in And~VI, $\langle P_{ab} \rangle = 0.588$d, is consistent 
with the galaxy's mean metallicity of $\langle {\rm [Fe/H]} \rangle = -1.58$ 
and follows the trend of the Galactic 
dwarf spheroidal galaxies filling in the gap between the Oosterhoff groups.  
The location of the RRab stars in a period-amplitude diagram is consistent 
with Galactic globular clusters of similar metallicity.  We were also able 
to show that a number of RR~Lyrae metallicity indicators, such as Eq.\ 2 
from Alcock et al.\ (2000) and the Sandage (1982a,b) period shift, give 
results consistent with the mean abundance ($\langle {\rm [Fe/H]} \rangle 
=-1.58\pm0.20$) derived by AJD99 from the red giant branch.  Indeed, based 
on the properties of its variable stars, the And~VI dSph is 
indistinguishable from the Galactic dSph companions.

\acknowledgements

This research was supported in part by NASA through grant number GO-08272 
from the Space Telescope Science Institute, which is operated by AURA, Inc.\, 
under NASA contract NAS 5-26555.

We would like to thank Dr.\ Peter Stetson for graciously sharing his PSFs for 
the WFPC2 and for the use of his data reduction programs.  Thanks to Dr. Andrew 
Layden for the use of his light curve analysis programs.  Thanks also to 
all of the groups who allowed us to use their results in this paper:  Karen 
Kinemuchi and collaborators, Dr.\ David Bersier, and Dr.\ Janusz Kaluzny.  
We would also like to thank Dr.\ Gisella Clementini for testing our period 
determinations.

\begin{deluxetable}{cccccccccccc} 
\tablewidth{0pc} 
\tabletypesize{\scriptsize} 
\tablecaption{Light Curve Properties \label{tbl-1}} 
\tablehead{
\colhead{ID} & \colhead{RA (2000)} & \colhead{Dec (2000)} & \colhead{Period} & 
\colhead{$\langle V \rangle$} & 
\colhead{$\langle B \rangle$} & \colhead{$(B-V)_{\rm mag}$}  & 
\colhead{$A_V$} & \colhead{$A_B$} & \colhead{$\Delta\log\,P$} & 
\colhead{[Fe/H]} & \colhead{Classification} \\ 
          } 
\startdata 
V01  & 23:51:47.9 & 24:34:39.5 & 0.579 & 25.387 & 25.734 & 0.389 & 0.99 & 1.40 & -0.03 & -1.81 & ab \\ 
V02  & 23:51:47.7 & 24:34:37.2 & 0.385 & 25.286 & 25.700 & 0.429 & 0.51 & 0.72 & \nodata & \nodata & c \\ 
V03  & 23:51:47.4 & 24:34:45.1 & 0.434 & 25.217 & 25.570 & 0.362 & 0.41 & 0.57 & \nodata & \nodata & c \\ 
V04  & 23:51:47.3 & 24:34:47.4 & 0.365 & 25.282 & 25.561 & 0.295 & 0.54 & 0.76 & \nodata & \nodata & c \\ 
V05  & 23:51:48.6 & 24:34:14.0 & 0.427 & 25.255 & 25.665 & 0.414 & 0.28 & 0.40 & \nodata & \nodata & c \\ 
V06  & 23:51:48.3 & 24:34:16.0 & 0.629 & 24.532 & 24.873 & 0.369 & 0.74 & 1.05 & \nodata & \nodata & AC \\ 
V07  & 23:51:48.7 & 24:34:02.4 & 0.652 & 25.330 & 25.655 & 0.335 & 0.48 & 0.68 &  0.01 & -1.59 & ab \\ 
V08  & 23:51:49.2 & 24:33:47.5 & 0.590 & 25.209 & 25.491 & 0.299 & 0.64 & 0.90 &  0.03 & -1.42 & ab \\ 
V09  & 23:51:47.6 & 24:34:27.8 & 0.576 & 25.321 & 25.627 & 0.356 & 1.07 & 1.51 & -0.04 & -1.90 & ab \\ 
V10  & 23:51:48.5 & 24:34:03.5 & 0.335 & 25.256 & 25.659 & 0.418 & 0.52 & 0.73 & \nodata & \nodata & c \\ 
V11  & 23:51:46.9 & 24:34:42.0 & 0.534 & 25.125 & 25.352 & 0.256 & 0.82 & 1.17 &  0.03 & -1.28 & ab \\ 
V12  & 23:51:47.4 & 24:34:27.8 & 0.549 & 25.329 & 25.711 & 0.417 & 0.92 & 1.30 &  0.00 & -1.52 & ab \\ 
V13  & 23:51:48.5 & 24:33:57.0 & 0.395 & 25.132 & 25.385 & 0.266 & 0.46 & 0.65 & \nodata & \nodata & c \\ 
V14  & 23:51:47.3 & 24:34:27.5 & 0.548 & 25.341 & 25.590 & 0.302 & 1.03 & 1.46 & -0.02 & -1.66 & ab \\ 
V15  & 23:51:46.8 & 24:34:34.2 & 0.624 & 25.353 & 25.644 & 0.323 & 0.87 & 1.23 & -0.04 & -1.94 & ab \\ 
V16  & 23:51:47.8 & 24:34:04.6 & 0.535 & 25.215 & 25.538 & 0.377 & 1.05 & 1.49 & -0.01 & -1.59 & ab \\ 
V17  & 23:51:47.5 & 24:34:12.4 & 0.555 & 25.306 & 25.611 & 0.326 & 0.65 & 0.92 &  0.05 & -1.20 & ab \\ 
V18  & 23:51:48.1 & 24:33:50.4 & 0.532 & 25.119 & 25.421 & 0.341 & 0.96 & 1.36 &  0.01 & -1.45 & ab \\ 
V19  & 23:51:48.0 & 24:33:51.8 & 0.605 & 25.275 & 25.644 & 0.390 & 0.69 & 0.98 &  0.00 & -1.58 & ab \\ 
V20  & 23:51:47.3 & 24:34:03.5 & 0.591 & 25.267 & 25.656 & 0.415 & 0.73 & 1.03 &  0.01 & -1.55 & ab \\ 
V21  & 23:51:46.5 & 24:34:14.2 & 0.490 & 25.102 & 25.374 & 0.332 & 1.14 & 1.61 &  0.01 & -1.37 & ab \\ 
V22  & 23:51:46.0 & 24:34:25.0 & 0.519 & 25.130 & 25.400 & 0.308 & 0.94 & 1.34 &  0.02 & -1.33 & ab \\ 
V23  & 23:51:45.6 & 24:34:26.6 & 0.673 & 25.304 & 25.673 & 0.383 & 0.58 & 0.82 & -0.02 & -1.85 & ab \\ 
V24  & 23:51:45.6 & 24:34:26.3 & 0.593 & 25.216 & 25.572 & 0.370 & 0.57 & 0.81 &  0.03 & -1.35 & ab \\ 
V25  & 23:51:47.1 & 24:33:46.3 & 0.626 & 25.338 & 25.710 & 0.397 & 0.78 & 1.11 & -0.03 & -1.84 & ab \\ 
V26  & 23:51:46.6 & 24:33:56.4 & 0.512 & 25.306 & 25.658 & 0.417 & 1.12 & 1.58 &  0.00 & -1.51 & ab \\ 
V27  & 23:51:46.4 & 24:33:56.6 & 0.599 & 25.284 & 25.568 & 0.319 & 0.90 & 1.27 & -0.03 & -1.82 & ab \\ 
V28  & 23:51:47.0 & 24:33:36.0 & 0.433 & 25.216 & 25.510 & 0.303 & 0.40 & 0.56 & \nodata & \nodata & c \\ 
V29  & 23:51:45.6 & 24:34:09.0 & 0.536 & 25.104 & 25.356 & 0.297 & 1.02 & 1.44 &  0.00 & -1.56 & ab \\ 
V30  & 23:51:45.6 & 24:34:09.0 & 0.319 & 25.354 & 25.716 & 0.380 & 0.56 & 0.79 & \nodata & \nodata & c \\ 
V31  & 23:51:45.9 & 24:34:01.2 & 0.647 & 25.161 & 25.567 & 0.423 & 0.59 & 0.83 & -0.01 & -1.71 & ab \\ 
V32  & 23:51:44.4 & 24:34:29.7 & 0.717 & 25.224 & 25.522 & 0.309 & 0.49 & 0.69 & -0.03 & -1.97 & ab \\ 
V33  & 23:51:45.1 & 24:34:06.7 & 0.567 & 25.183 & 25.494 & 0.341 & 0.83 & 1.18 &  0.01 & -1.52 & ab \\ 
V34  & 23:51:45.0 & 24:34:04.5 & 0.651 & 25.270 & 25.631 & 0.377 & 0.56 & 0.82 & \nodata & \nodata & Contact Binary? \\ 
V35  & 23:51:46.5 & 24:33:26.5 & 0.604 & 25.237 & 25.595 & 0.385 & 0.81 & 1.14 & -0.02 & -1.74 & ab \\ 
V36  & 23:51:45.6 & 24:33:47.1 & 0.539 & 25.406 & 25.682 & 0.331 & 1.12 & 1.58 & -0.02 & -1.71 & ab \\ 
V37  & 23:51:45.0 & 24:34:01.7 & 0.691 & 25.362 & 25.674 & 0.330 & 0.64 & 0.91 & -0.04 & -2.03 & ab \\ 
V38  & 23:51:45.7 & 24:33:31.3 & 0.548 & 25.143 & 25.507 & 0.402 & 0.87 & 1.23 &  0.01 & -1.44 & ab \\ 
V39  & 23:51:43.6 & 24:34:22.2 & 0.610 & 25.397 & 25.665 & 0.291 & 0.75 & 1.06 & -0.01 & -1.56 & ab \\ 
V40  & 23:51:43.4 & 24:34:21.5 & 0.434 & 25.197 & 25.449 & 0.266 & 0.50 & 0.71 & \nodata & \nodata & c \\ 
V41  & 23:51:43.5 & 24:34:15.1 & 0.512 & 25.145 & 25.466 & 0.376 & 1.12 & 1.59 &  0.00 & -1.51 & ab \\ 
V42  & 23:51:44.0 & 24:34:01.4 & 0.596 & 25.235 & 25.589 & 0.388 & 0.82 & 1.15 & -0.01 & -1.70 & ab \\ 
V43  & 23:51:43.7 & 24:33:50.4 & 0.587 & 25.334 & 25.780 & 0.463 & 0.65 & 0.92 &  0.02 & -1.42 & ab \\ 
V44  & 23:51:47.5 & 24:34:49.4 & 0.760 & 23.624 & 23.988 & 0.378 & 0.53 & 0.75 & \nodata & \nodata & AC \\ 
V45  & 23:51:43.5 & 24:34:32.8 & 0.546 & 25.340 & 25.553 & 0.266 & 1.09 & 1.55 & -0.03 & -1.72 & ab \\ 
V46  & 23:51:43.7 & 24:34:36.3 & 0.652 & 25.307 & 25.546 & 0.251 & 0.49 & 0.70 &  0.01 & -1.61 & ab \\ 
V47  & 23:51:46.2 & 24:34:57.6 & 0.570 & 25.272 & 25.551 & 0.325 & 0.97 & 1.38 & -0.02 & -1.73 & ab \\ 
V48  & 23:51:45.4 & 24:34:51.5 & 0.324 & 25.342 & 25.645 & 0.316 & 0.48 & 0.68 & \nodata & \nodata & c \\ 
V49  & 23:51:46.2 & 24:34:59.1 & 0.545 & 25.340 & 25.677 & 0.411 & 1.20 & 1.60 & -0.03 & -1.86 & ab \\ 
V50  & 23:51:45.5 & 24:34:56.5 & 0.614 & 25.364 & 25.717 & 0.363 & 0.52 & 0.73 &  0.03 & -1.42 & ab \\ 
V51  & 23:51:45.8 & 24:35:00.1 & 0.607 & 25.431 & 25.780 & 0.381 & 0.85 & 1.20 & -0.03 & -1.81 & ab \\ 
V52  & 23:51:45.9 & 24:35:01.2 & 0.725 & 23.570 & 23.884 & 0.330 & 0.52 & 0.74 & \nodata & \nodata & AC \\ 
V53  & 23:51:43.3 & 24:34:42.7 & 0.585 & 25.223 & 25.503 & 0.302 & 0.72 & 1.02 &  0.01 & -1.50 & ab \\ 
V54  & 23:51:42.5 & 24:34:42.3 & 0.518 & 25.382 & 25.603 & 0.257 & 0.82 & 1.17 &  0.05 & -1.16 & ab \\ 
V55  & 23:51:46.2 & 24:35:09.7 & 0.586 & 25.320 & 25.673 & 0.381 & 0.77 & 1.09 &  0.00 & -1.57 & ab \\ 
V56  & 23:51:45.0 & 24:35:03.2 & 0.556 & 25.367 & 25.694 & 0.368 & 0.89 & 1.26 &  0.00 & -1.53 & ab \\ 
V57  & 23:51:44.0 & 24:34:56.5 & 0.573 & 25.342 & 25.683 & 0.369 & 0.77 & 1.09 &  0.01 & -1.48 & ab \\ 
V58  & 23:51:46.0 & 24:35:12.2 & 0.654 & 25.363 & 25.651 & 0.301 & 0.52 & 0.74 &  0.00 & -1.66 & ab \\ 
V59  & 23:51:44.7 & 24:35:06.9 & 0.636 & 25.312 & 25.624 & 0.325 & 0.50 & 0.71 &  0.02 & -1.52 & ab \\ 
V60  & 23:51:44.2 & 24:35:05.6 & 0.590 & 25.398 & 25.803 & 0.433 & 0.81 & 1.15 & -0.01 & -1.65 & ab \\ 
V61  & 23:51:44.2 & 24:35:12.3 & 0.527 & 25.487 & 25.846 & 0.394 & 0.87 & 1.22 &  0.03 & -1.29 & ab \\ 
V62  & 23:51:41.7 & 24:34:56.4 & 0.644 & 25.235 & 25.551 & 0.343 & 0.80 & 1.13 & -0.04 & -1.97 & ab \\ 
V63  & 23:51:45.6 & 24:35:25.3 & 0.651 & 25.290 & 25.595 & 0.342 & 0.93 & 1.32 & -0.07 & -2.18 & ab \\ 
V64  & 23:51:42.6 & 24:35:07.3 & 0.541 & 25.208 & 25.471 & 0.280 & 0.57 & 0.81 &  0.07 & -1.00 & ab \\ 
V65  & 23:51:43.2 & 24:35:15.1 & 0.412 & 25.271 & 25.538 & 0.276 & 0.40 & 0.57 & \nodata & \nodata & c \\ 
V66  & 23:51:44.5 & 24:35:25.8 & 0.633 & 25.383 & 25.708 & 0.360 & 0.91 & 1.24 & -0.01 & -1.77 & ab \\ 
V67  & 23:51:41.1 & 24:35:01.4 & 0.415 & 25.161 & 25.523 & 0.368 & 0.32 & 0.45 & \nodata & \nodata & c \\ 
V68  & 23:51:44.8 & 24:35:30.7 & 0.594 & 25.103 & 25.408 & 0.330 & 0.72 & 1.03 &  0.01 & -1.55 & ab \\ 
V69  & 23:51:43.7 & 24:35:24.4 & 0.353 & 25.239 & 25.606 & 0.385 & 0.56 & 0.79 & \nodata & \nodata & c \\ 
V70  & 23:51:43.2 & 24:35:21.3 & 0.529 & 25.296 & 25.681 & 0.442 & 1.08 & 1.53 & -0.01 & -1.59 & ab \\ 
V71  & 23:51:41.3 & 24:35:10.8 & 0.546 & 25.300 & 25.720 & 0.451 & 0.79 & 1.12 &  0.03 & -1.32 & ab \\ 
V72  & 23:51:45.3 & 24:35:40.9 & 0.578 & 25.468 & 25.720 & 0.274 & 0.68 & 0.96 &  0.03 & -1.40 & ab \\ 
V73  & 23:51:43.9 & 24:35:35.6 & 0.660 & 25.321 & 25.578 & 0.269 & 0.52 & 0.74 &  0.00 & -1.69 & ab \\ 
V74  & 23:51:44.6 & 24:35:43.2 & 0.562 & 25.273 & 25.532 & 0.328 & 1.14 & 1.62 & -0.01 & -1.58 & ab \\ 
V75  & 23:51:44.3 & 24:35:45.2 & 0.599 & 25.355 & 25.760 & 0.434 & 0.83 & 1.17 & -0.02 & -1.73 & ab \\ 
V76  & 23:51:43.5 & 24:35:39.8 & 0.528 & 25.257 & 25.523 & 0.298 & 0.80 & 1.13 &  0.04 & -1.21 & ab \\ 
V77  & 23:51:41.0 & 24:35:22.7 & 0.569 & 25.395 & 25.738 & 0.364 & 0.72 & 1.02 &  0.03 & -1.39 & ab \\ 
V78  & 23:51:43.6 & 24:35:42.9 & 0.480 & 25.331 & 25.738 & 0.450 & 1.01 & 1.43 &  0.05 & -1.12 & ab \\ 
V79  & 23:51:45.0 & 24:35:52.7 & 0.639 & 25.399 & 25.704 & 0.313 & 0.43 & 0.61 &  0.03 & -1.45 & ab \\ 
V80  & 23:51:43.7 & 24:35:44.9 & 0.381 & 25.354 & 25.635 & 0.295 & 0.47 & 0.67 & \nodata & \nodata & c \\ 
V81  & 23:51:43.7 & 24:35:45.3 & 0.386 & 25.361 & 25.628 & 0.285 & 0.56 & 0.79 & \nodata & \nodata & c \\ 
V82  & 23:51:43.8 & 24:35:48.3 & 0.554 & 25.372 & 25.804 & 0.447 & 0.60 & 0.84 &  0.06 & -1.13 & ab \\ 
V83  & 23:51:43.6 & 24:35:47.7 & 0.674 & 23.471 & 23.789 & 0.332 & 0.50 & 0.70 & \nodata & \nodata & AC \\ 
V84  & 23:51:47.2 & 24:35:10.6 & 1.357 & 23.661 & 24.054 & 0.408 & 0.60 & 0.84 & \nodata & \nodata & AC \\ 
V85  & 23:51:45.4 & 24:35:59.9 & 0.651 & 25.243 & 25.683 & 0.457 & 0.64 & 0.90 & -0.02 & -1.80 & ab \\ 
V86  & 23:51:47.3 & 24:35:10.3 & 0.307 & 25.254 & 25.683 & 0.436 & 0.36 & 0.51 & \nodata & \nodata & c \\ 
V87  & 23:51:47.8 & 24:35:01.5 & 0.599 & 25.372 & 25.764 & 0.417 & 0.75 & 1.05 &  0.00 & -1.63 & ab \\ 
V88  & 23:51:46.4 & 24:35:55.1 & 0.621 & 25.383 & 25.651 & 0.282 & 0.58 & 0.82 &  0.01 & -1.54 & ab \\ 
V89  & 23:51:47.9 & 24:35:17.9 & 0.627 & 25.360 & 25.690 & 0.351 & 0.67 & 0.94 & -0.01 & -1.70 & ab \\ 
V90  & 23:51:46.9 & 24:35:44.8 & 0.415 & 25.119 & 25.378 & 0.282 & 0.63 & 0.89 & \nodata & \nodata & c \\ 
V91  & 23:51:46.8 & 24:35:50.0 & 0.349 & 25.364 & 25.601 & 0.255 & 0.55 & 0.78 & \nodata & \nodata & c \\ 
V92  & 23:51:46.1 & 24:36:09.7 & 0.661 & 25.364 & 25.737 & 0.391 & 0.58 & 0.81 & -0.01 & -1.78 & ab \\ 
V93  & 23:51:47.8 & 24:35:33.8 & 0.477 & 24.749 & 25.130 & 0.402 & 0.60 & 0.84 & \nodata & \nodata & AC \\ 
V94  & 23:51:47.9 & 24:35:33.0 & 0.579 & 25.292 & 25.675 & 0.411 & 0.75 & 1.06 &  0.01 & -1.50 & ab \\ 
V95  & 23:51:47.5 & 24:35:43.7 & 0.548 & 25.284 & 25.629 & 0.389 & 0.93 & 1.31 &  0.05 & -1.52 & ab \\ 
V96  & 23:51:48.7 & 24:35:22.7 & 0.674 & 25.204 & 25.634 & 0.445 & 0.56 & 0.79 & -0.02 & -1.83 & ab \\ 
V97  & 23:51:48.3 & 24:35:37.8 & 0.614 & 25.371 & 25.658 & 0.326 & 0.96 & 1.37 & -0.05 & -2.01 & ab \\ 
V98  & 23:51:48.6 & 24:35:31.8 & 0.563 & 25.344 & 25.710 & 0.385 & 0.67 & 0.95 &  0.04 & -1.28 & ab \\ 
V99  & 23:51:48.1 & 24:35:46.0 & 0.608 & 25.358 & 25.774 & 0.443 & 0.80 & 1.13 & -0.02 & -1.75 & ab \\ 
V100 & 23:51:48.3 & 24:35:40.6 & 0.666 & 25.215 & 25.588 & 0.386 & 0.54 & 0.77 & -0.01 & -1.75 & ab \\ 
V101 & 23:51:49.2 & 24:35:23.0 & 0.595 & 25.292 & 25.675 & 0.402 & 0.62 & 0.88 &  0.02 & -1.43 & ab \\ 
V102 & 23:51:47.8 & 24:36:03.7 & 0.621 & 25.371 & 25.755 & 0.401 & 0.63 & 0.88 &  0.01 & -1.61 & ab \\ 
V103 & 23:51:47.6 & 24:36:15.6 & 0.734 & \nodata & \nodata & \nodata & \nodata & \nodata & \nodata & \nodata & ab \\ 
V104 & 23:51:47.6 & 24:36:16.2 & 0.524 & 25.299 & 25.635 & 0.373 & 0.87 & 1.24 &  0.03 & -1.27 & ab \\ 
V105 & 23:51:47.7 & 24:36:15.6 & 0.524 & 25.141 & 25.415 & 0.322 & 0.95 & 1.35 &  0.02 & -1.38 & ab \\ 
V106 & 23:51:49.9 & 24:35:26.0 & 0.544 & 25.312 & 25.745 & 0.469 & 0.88 & 1.24 &  0.02 & -1.43 & ab \\ 
V107 & 23:51:50.1 & 24:35:22.2 & 0.590 & 25.457 & 25.752 & 0.335 & 0.88 & 1.25 & -0.02 & -1.74 & ab \\ 
V108 & 23:51:48.1 & 24:36:19.4 & 0.572 & 25.339 & 25.714 & 0.403 & 0.74 & 1.05 &  0.02 & -1.44 & ab \\ 
V109 & 23:51:49.6 & 24:35:42.9 & 0.406 & 25.237 & 25.623 & 0.396 & 0.42 & 0.59 & \nodata & \nodata & c \\ 
V110 & 23:51:49.1 & 24:36:04.1 & 0.636 & 25.313 & 25.715 & 0.414 & 0.51 & 0.73 &  0.01 & -1.54 & ab \\ 
V111 & 23:51:49.4 & 24:36:02.6 & 0.698 & 25.284 & 25.592 & 0.319 & 0.51 & 0.72 & -0.02 & -1.90 & ab \\ 
V112 & 23:51:50.5 & 24:35:36.8 & 0.597 & 25.276 & 25.655 & 0.405 & 0.80 & 1.13 & -0.01 & -1.68 & ab \\ 
V113 & 23:51:49.1 & 24:36:17.6 & 0.509 & 25.374 & 25.792 & 0.454 & 0.90 & 1.27 &  0.04 & -1.20 & ab \\ 
V114 & 23:51:49.7 & 24:36:03.7 & 0.619 & 25.207 & 25.523 & 0.343 & 0.73 & 1.03 & -0.01 & -1.73 & ab \\ 
V115 & 23:51:49.9 & 24:36:05.7 & 0.572 & 25.360 & 25.719 & 0.377 & 0.67 & 0.94 &  0.03 & -1.34 & ab \\ 
V116 & 23:51:50.6 & 24:35:53.7 & 0.583 & 25.209 & 25.490 & 0.312 & 0.85 & 1.20 & -0.01 & -1.65 & ab \\ 
V117 & 23:51:51.5 & 24:35:45.5 & 0.366 & 25.226 & 25.631 & 0.417 & 0.65 & 0.46 & \nodata & \nodata & c \\ 
V118 & 23:51:50.6 & 24:36:09.5 & 0.511 & 25.203 & 25.476 & 0.293 & 0.71 & 1.00 &  0.07 & -0.96 & ab \\ 
\enddata
\end{deluxetable}

\clearpage 

\begin{deluxetable}{cccccc} 
\tablewidth{0pt} 
\footnotesize 
\tablecaption{Photometry of the Variable Stars ($B$)\label{tbl-2}}
\tablehead{
\colhead{} & \multicolumn{2}{c}{V01} & & \multicolumn{2}{c}{V02} \\ 
\cline{2-3} \cline{5-6} \\ 
\colhead{HJD-2451000} & \colhead{$B$} & \colhead{$\sigma_{B}$} & & 
\colhead{$B$} & \colhead{$\sigma_{B}$} 
          }
\startdata 
476.729 &  26.078  &  0.208 &&  26.150  &  0.342 \\ 
476.746 &  25.592  &  0.214 &&  25.973  &  0.213 \\ 
476.797 &  24.883  &  0.100 &&  25.524  &  0.113 \\ 
476.813 &  25.130  &  0.156 &&  25.453  &  0.238 \\ 
476.863 &  25.418  &  0.235 &&  \nodata  &  \nodata \\ 
476.880 &  25.498  &  0.164 &&  25.273  &  0.123 \\ 
476.931 &  25.654  &  0.223 &&  25.545  &  0.146 \\ 
476.947 &  25.734  &  0.179 &&  25.687  &  0.112 \\ 
479.011 &  26.249  &  0.210 &&  25.861  &  0.138 \\ 
479.028 &  26.243  &  0.236 &&  26.161  &  0.195 \\ 
479.078 &  25.126  &  0.145 &&  25.784  &  0.134 \\ 
479.095 &  25.099  &  0.193 &&  25.658  &  0.115 \\ 
479.146 &  25.227  &  0.094 &&  25.491  &  0.106 \\ 
479.162 &  25.253  &  0.207 &&  25.214  &  0.113 \\ 
479.213 &  25.553  &  0.143 &&  \nodata  &  \nodata \\ 
479.230 &  25.627  &  0.175 &&  25.458  &  0.152 \\
\enddata 
\tablecomments{The complete version of this table is in the electronic 
edition of the Journal.  The printed edition contains only a sample.}
\end{deluxetable}

\begin{deluxetable}{cccccc} 
\tablewidth{0pt} 
\footnotesize 
\tablecaption{Photometry of the Variable Stars ($V$)\label{tbl-3}}
\tablehead{
\colhead{} & \multicolumn{2}{c}{V01} & & \multicolumn{2}{c}{V02} \\ 
\cline{2-3} \cline{5-6} \\ 
\colhead{HJD-2451000} & \colhead{$V$} & \colhead{$\sigma_{V}$} & & 
\colhead{$V$} & \colhead{$\sigma_{V}$} 
          }
\startdata 
476.597 &  25.763  &  0.085 &&  25.316  &  0.101 \\ 
476.613 &  25.767  &  0.206 &&  25.451  &  0.075 \\ 
476.661 &  25.772  &  0.208 &&  25.450  &  0.102 \\ 
476.677 &  25.658  &  0.197 &&  25.539  &  0.129 \\ 
478.879 &  25.557  &  0.087 &&  25.332  &  0.131 \\ 
478.895 &  25.789  &  0.120 &&  25.508  &  0.173 \\ 
478.943 &  25.580  &  0.100 &&  25.392  &  0.154 \\ 
478.960 &  25.511  &  0.151 &&  25.576  &  0.127 \\ 
\enddata 
\tablecomments{The complete version of this table is in the electronic 
edition of the Journal.  The printed edition contains only a sample.}
\end{deluxetable}

\clearpage 

\begin{deluxetable}{lccrllccrrccl} 
\tablewidth{0pc} 
\tabletypesize{\scriptsize}
\tablecaption{Properties of Anomalous Cepheids\label{tbl-4}}
\tablehead{
\colhead{System} & \colhead{$(m-M)_0$} & \colhead{E(\bv)} & 
\colhead{ID} & \colhead{Mode} & \colhead{Period} & 
\colhead{$\langle V \rangle$} & \colhead{$\langle B \rangle$} & 
\colhead{$M_V$} & \colhead{$M_B$} & \colhead{$A_V$} & 
\colhead{$A_B$} & \colhead{References} 
          }
\startdata 
And VI & 24.45 & 0.06 &   6 & F & 0.629 & 24.53 & 24.87 & --0.10 &   0.18 & 0.74 & 1.05 & AJD99 \\ 
       &       &      &  44 & H & 0.760 & 23.62 & 23.99 & --1.01 & --0.71 & 0.53 & 0.75 \\ 
       &       &      &  52 & H & 0.725 & 23.57 & 23.88 & --1.07 & --0.81 & 0.52 & 0.74 \\ 
       &       &      &  83 & H & 0.674 & 23.47 & 23.79 & --1.17 & --0.91 & 0.50 & 0.70 \\ 
       &       &      &  84 & F & 1.357 & 23.66 & 24.05 & --0.98 & --0.64 & 0.60 & 0.84 \\ 
       &       &      &  93 & F & 0.477 & 24.75 & 25.13 &   0.11 &   0.43 & 0.60 & 0.84 \\
Leo I  & 21.93 & 0.01 &   1 & F & 1.322 & \nodata & 20.7 & \nodata & --1.3 & \nodata & 1.6 & HW78 \\ 
       &       &      &   2 & F & 1.824 & \nodata & 21.0 & \nodata & --1.0 & \nodata & 1.5 \\ 
       &       &      &   8 & F & 2.374 & \nodata & 20.1 & \nodata & --1.9 & \nodata & 1.0 \\ 
       &       &      &  10 & F & 2.301 & \nodata & 21.0 & \nodata & --1.0 & \nodata & 2.0 \\ 
       &       &      &  11 & H & 0.851 & \nodata & 21.2 & \nodata & --0.8 & \nodata & 0.7 \\ 
       &       &      &  13 & F & 0.956 & \nodata & 21.8 & \nodata & --0.2 & \nodata & 1.1 \\ 
       &       &      &  15 & F & 1.024 & \nodata & 21.6 & \nodata & --0.4 & \nodata & 1.2 \\ 
       &       &      &  16 & F & 1.499 & \nodata & 20.9 & \nodata & --1.1 & \nodata & 2.3 \\ 
       &       &      &  17 & H & 0.799 & \nodata & 20.6 & \nodata & --1.4 & \nodata & 1.5 \\ 
       &       &      &  19 & F & 1.629 & \nodata & 20.7 & \nodata & --1.3 & \nodata & 1.0 \\ 
       &       &      &  20 & F & 1.522 & \nodata & 21.3 & \nodata & --0.7 & \nodata & 1.0 \\ 
       &       &      &  23 & F & 1.100 & \nodata & 21.4 & \nodata & --0.6 & \nodata & 0.7 \\ 
Leo II & 21.59 & 0.02 &   1* & F? & 0.408 & 21.97 & \nodata &   0.31 & \nodata & 0.76 & \nodata & SM00 \\
       &       &      &  27  & F  & 1.486 & 20.45 & \nodata & --1.20 & \nodata & 1.24 & \nodata \\ 
       &       &      &  51* & H  & 0.396 & 21.59 & \nodata & --0.06 & \nodata & 0.77 & \nodata \\ 
       &       &      & 203  & F  & 1.380 & 20.59 & \nodata & --1.06 & \nodata & 1.05 & \nodata \\ 
Draco  & 19.49 & 0.03 & 055 & F & 0.552 & 19.44 & \nodata & --0.14 & \nodata & 0.57 & \nodata & ZS76; K02 \\ 
       &       &      & 119 & F & 0.907 & 19.03 & \nodata & --0.55 & \nodata & 1.00 & \nodata \\ 
       &       &      & 134 & H & 0.592 & 18.78 & 19.06 & --0.79 & --0.54 & 0.88 & 1.13 \\ 
       &       &      & 141 & F & 0.901 & 19.12 & 19.43 & --0.46 & --0.17 & 0.72 & 1.15 \\ 
       &       &      & 157 & F & 0.936 & 18.77 & 19.24 & --0.80 & --0.36 & 1.04 & 1.35 \\ 
       &       &      & 194 & F & 1.590 & 18.12 & 18.53 & --1.45 & --1.07 & 0.46 & 0.52 \\ 
       &       &      & 204 & H & 0.454 & 19.24 & 19.49 & --0.33 & --0.11 & 0.78 & 1.02 \\ 
       &       &      & 208 & F & 0.608 & 19.28 & \nodata & --0.29 & \nodata & 0.33 & \nodata \\ 
Ursa Minor & 19.16 & 0.03 &  1  & F & 0.471 & \nodata & 19.70 & \nodata &   0.42 & \nodata & 1.05 & NWM88 \\ 
           &       &      &  6  & H & 0.724 & \nodata & 18.25 & \nodata & --1.03 & \nodata & 0.66 \\ 
           &       &      & 11* & F & 0.675 & \nodata & 19.29 & \nodata &   0.01 & \nodata & 1.59 \\ 
           &       &      & 56  & F & 0.611 & \nodata & 19.38 & \nodata &   0.10 & \nodata & 0.48 \\ 
           &       &      & 59  & H & 0.390 & \nodata & 19.56 & \nodata &   0.28 & \nodata & 1.01 \\ 
           &       &      & 62* & H & 0.421 & \nodata & 19.33 & \nodata &   0.05 & \nodata & 1.17 \\ 
Carina & 20.14 & 0.04 &   1 & F & 0.611 & \nodata & 20.20 & \nodata & --0.10 & \nodata & 0.51 & SMS86 \\ 
       &       &      &  14 & H & 0.480 & \nodata & 20.11 & \nodata & --0.19 & \nodata & 0.89 \\ 
       &       &      &  27 & H & 0.511 & \nodata & 19.37 & \nodata & --0.93 & \nodata & 1.68 \\ 
       &       &      &  29 & H & 0.726 & \nodata & 19.20 & \nodata & --1.10 & \nodata & 1.19 \\ 
       &       &      &  33 & F? & 0.575 & \nodata & 20.16 & \nodata & --0.14 & \nodata & 0.91 \\ 
       &       &      & 129 & H & 0.640 & \nodata & 19.29 & \nodata & --1.01 & \nodata & 0.99 \\ 
       &       &      & 149 & H & 0.465 & \nodata & 20.31 & \nodata &   0.01 & \nodata & 1.30 \\ 
Sculptor & 19.56 & 0.02 &   26 & F & 1.346 & 18.55 & \nodata & --1.08 & \nodata & 0.80 & \nodata & K95 \\ 
         &       &      &  119 & F & 1.159 & 18.86 & \nodata & --0.76 & \nodata & 0.55 & \nodata \\ 
         &       &      & 5689 & F & 0.855 & 19.14 & \nodata & --0.47 & \nodata & 0.70 & \nodata \\ 
Fornax & 20.70 & 0.03 &   1 & F  & 0.785 & 20.38 & \nodata & --0.41 & \nodata & 0.95 & \nodata & LAZ86 \\ 
       &       &      & 825 & F  & 1.045 & 19.80 & \nodata & --0.99 & \nodata & \nodata & \nodata & BW02 \\ 
       &       &      & 012 & F  & 1.250 & 19.91 & \nodata & --0.89 & \nodata & \nodata & \nodata \\ 
       &       &      & 316 & F  & 0.508 & 20.79 & \nodata & --0.01 & \nodata & \nodata & \nodata \\ 
       &       &      & 122 & F  & 0.504 & 20.83 & \nodata &   0.04 & \nodata & \nodata & \nodata \\ 
       &       &      & 621 & F  & 0.546 & 20.65 & \nodata & --0.15 & \nodata & \nodata & \nodata \\ 
       &       &      & 125 & F  & 0.573 & 20.62 & \nodata & --0.18 & \nodata & \nodata & \nodata \\ 
       &       &      & 001 & F  & 0.922 & 20.57 & \nodata & --0.23 & \nodata & \nodata & \nodata \\ 
       &       &      & 340 & F  & 1.311 & 20.25 & \nodata & --0.54 & \nodata & \nodata & \nodata \\ 
       &       &      & 433 & F  & 0.611 & 20.85 & \nodata &   0.05 & \nodata & \nodata & \nodata \\ 
       &       &      & 601 & F  & 0.574 & 21.02 & \nodata &   0.23 & \nodata & \nodata & \nodata \\ 
       &       &      & 846 & H? & 0.416 & 20.83 & \nodata &   0.04 & \nodata & \nodata & \nodata \\ 
       &       &      & 928 & H  & 0.533 & 20.36 & \nodata & --0.43 & \nodata & \nodata & \nodata \\ 
       &       &      & 641 & F  & 0.533 & 20.84 & \nodata &   0.05 & \nodata & \nodata & \nodata \\ 
       &       &      & 024 & F  & 1.198 & 19.99 & \nodata & --0.80 & \nodata & \nodata & \nodata \\ 
       &       &      & 802 & F  & 0.838 & 20.48 & \nodata & --0.31 & \nodata & \nodata & \nodata \\ 
       &       &      & 335 & F  & 0.506 & 20.83 & \nodata &   0.03 & \nodata & \nodata & \nodata \\ 
       &       &      & 552 & F  & 0.481 & 20.79 & \nodata & --0.01 & \nodata & \nodata & \nodata \\   
Sextans & 19.74 & 0.03 & 1  & H  & 0.693 & 18.83 & 19.08 & --1.00 & --0.79 & 0.83 & 1.11 & MFK95 \\ 
        &       &      & 5  & F  & 0.861 & 19.54 & 19.85 & --0.30 & --0.10 & 0.79 & 1.18 \\ 
        &       &      & 6  & F  & 0.922 & 19.19 & 19.46 & --0.65 & --0.40 & 1.31 & 1.64 \\ 
        &       &      & 9* & F? & 0.416 & 20.01 & 20.30 &   0.18 &   0.44 & 0.72 & 0.82 \\ 
NGC 5466 & 16.03 & 0.00 & 19 & H & 0.822 & 14.731 & 14.909 & --1.30 & --1.12 & 0.60 & 0.70 & CCN99 \\ 
\enddata
\tablecomments{AJD99 = Armandroff, Jacoby \& Davies (1999); HW78 = Hodge \& 
Wright (1978); SM00 = Siegel \& Majewski (2000); ZS76 = Zinn \& Searle (1976); 
K02 = Kinemuchi et al. (2002); NWM88 = Nemec, Wehlau, \& Mendes~de~Oliveira 
(1988); SMS86 = Saha, Monet, \& Seitzer (1986); K95 = Kaluzny et al. (1995); 
LAZ86 = Light, Armandroff, \& Zinn (1986); BW02 = Bersier \& Wood (2002); 
MFK95 = Mateo, Fischer, \& Krzeminski (1995); CCN99 = Corwin, Carney, \& 
Nifong (1999)} 
\end{deluxetable}

\clearpage 

\begin{deluxetable}{lc} 
\tablewidth{0pc} 
\footnotesize 
\tablecaption{Period--Amplitude Fits for the RR~Lyrae \label{tbl-5}} 
\tablehead{
\colhead{System} & \colhead{Amplitude ($V$) Equation} \\ 
          } 
\startdata 
Andromeda VI & $-7.036(\pm 0.719)\log\,P - 0.860(\pm 0.170)$ \\ 
Sculptor     & $-5.885(\pm 0.468)\log\,P - 0.544(\pm 0.109)$ \\ 
Leo II       & $-10.420(\pm 1.178)\log\,P - 1.445(\pm 0.253)$ \\ 
Draco        & $-9.328(\pm 0.738)\log\,P - 1.222(\pm 0.159)$ \\ 
\enddata 
\end{deluxetable}

\begin{deluxetable}{rcccccccl} 
\tablewidth{0pc} 
\tabletypesize{\scriptsize}
\tablecaption{Properties of RR~Lyrae in Various Systems \label{tbl-6}} 
\tablehead{
& \colhead{System} & \colhead{[Fe/H]} & \colhead{$\langle P_{ab} \rangle$} & 
\colhead{$\langle P_c \rangle$} & \colhead{$N_{ab}$}  & 
\colhead{$N_c$} & \colhead{$N_c/N_{RR}$} & \colhead{Source} \\ 
          } 
\startdata 
dSphs      & Ursa Minor   & -2.2 & 0.638 & 0.375 &  47 & 35  & 0.43 & Nemec, Wehlau, \& Mendes de Oliveira 1988 \\ 
           & Carina       & -2.0 & 0.620 & 0.348 &  49 &  9  & 0.16 & Saha, Monet, \& Seitzer 1986 \\ 
           & Draco        & -2.0 & 0.615 & 0.372 & 209 & 28  & 0.12 & Kinemuchi et al.\ 2002 \\ 
           & Leo II       & -1.9 & 0.619 & 0.363 &  92 & 30  & 0.25 & Siegel \& Majewski 2000 \\ 
           & Sculptor     & -1.8 & 0.586 & 0.336 & 134 & 88  & 0.40 & Kaluzny et al.\ 1995 \\ 
           & Sextans      & -1.7 & 0.606 & 0.355 &  26 &  7  & 0.21 & Mateo, Fischer, \& Krzeminski 1995 \\ 
           & Leo I        & -1.7 & 0.602 & \nodata & 63 & 11 & 0.15 & Held et al.\ 2001 \\ 
           & Andromeda VI & -1.6 & 0.588 & 0.382 &  90 & 20  & 0.18 & This Paper\\ 
           & Fornax       & -1.3 & 0.585 & 0.349 & 396 & 119 & 0.23 & Bersier \& Wood 2002\\ 
           & Sagittarius  & -1.0 & 0.574 & 0.322 & \nodata & \nodata & \nodata & Cseresnjes 2001 \\
LMC GCs    & NGC 1841     & -2.2 & 0.676 & 0.344 &  17 &  5  & 0.23 & Kinman, Stryker, \& Hesser 1976; Walker 1990 \\ 
           & NGC 2210     & -1.9 & 0.598 & 0.379 &  20 &  9  & 0.31 & Reid \& Freedman 1994 \\ 
           & NGC 1466     & -1.9 & 0.589 & 0.345 &  19 & 13  & 0.41 & Walker 1992b \\ 
           & NGC 1835     & -1.8 & 0.598 & 0.326 &  18 & 15  & 0.46 & Graham \& Ruiz 1974; Walker 1993 \\ 
           & NGC 2257     & -1.8 & 0.578 & 0.343 &  13 & 13  & 0.50 & Walker 1989 \\ 
           & GLC 0435-59  & -1.7 & 0.559 & 0.340 &  16 &  7  & 0.30 & Walker 1992a \\ 
Oosterhoff & Type I       & \nodata & 0.55 & 0.32 & \nodata & \nodata & 0.17 & Smith 1995 \\ 
           & Type II      & \nodata & 0.64 & 0.37 & \nodata & \nodata & 0.44 & Smith 1995 \\ 
\enddata 
\end{deluxetable}

\end{document}